\documentclass[12pt]{article}

\usepackage[margin=1.25in]{geometry}
\usepackage{graphicx}
\usepackage{booktabs}
\usepackage{xcolor, soul}
\usepackage{amsmath}
\usepackage{float}
\usepackage[natbib,sorting=none,style=numeric-comp]{biblatex}
\usepackage[colorlinks=true,citecolor=blue,linkcolor=blue,urlcolor=blue]{hyperref}
\usepackage{setspace}

\newcommand\secsep{\vspace{1.5em}\hrule\vspace{0.5em}}

\begin{document}

\begin{refsection}

\rule{15.24cm}{1mm}
\begin{center}
  {\LARGE\bfseries BaCd$_2$P$_2$: a promising impurity-tolerant counterpart of GaAs for photovoltaics}\\[1em]
  %{\LARGE\bfseries BaCd$_2$P$_2$: a promising defect-tolerant alternative to GaAs for photovoltaics}\\[1em]
  {\large
    Gideon Kassa\textsuperscript{1}, 
    Zhenkun Yuan\textsuperscript{1}, 
    Muhammad R.\ Hasan\textsuperscript{2},\\
    Guillermo L.\ Esparza\textsuperscript{3}, 
    David P.\ Fenning\textsuperscript{3}, 
    Geoffroy Hautier\textsuperscript{1,4,5},\\
    Kirill Kovnir\textsuperscript{2, 6}, 
    Jifeng Liu\textsuperscript{1}\footnotemark[2]
  }\\[1em]
  \textsuperscript{1}Thayer School of Engineering, Dartmouth College, Hanover, NH 03755, USA\\
  \textsuperscript{2}Department of Chemistry, Iowa State University, Ames, IA 50011, USA\\
  \textsuperscript{3}Aiiso Yufeng Li Family Department of Chemical \& Nanoengineering, UC San Diego, La Jolla, CA 92093, USA\\
  \textsuperscript{4}Department of Materials Science and NanoEngineering, Rice University, Houston, TX 77005, USA \\
  \textsuperscript{5}Rice Advanced Materials Institute, Rice University, Houston, TX 77005, USA\\
  \textsuperscript{6}Ames National Laboratory, U.S. Department of Energy, Ames, IA 50011, USA\\
  [0.5em]
\end{center}

%\footnotetext[1]{Code available on \href{github.com/gideon116/BaCd2P2-aPIT-counterpart-of-GaAs-for-PVs}{https://github.com/gideon116/BaCd2P2-aPIT-counterpart-of-GaAs-for-PVs} }
\footnotetext[2]{Corresponding author. Email: \texttt{Jifeng.Liu@dartmouth.edu}}

%–– Abstract ––
\secsep
\begin{abstract}
\noindent\small

$\mathrm{BaCd_2P_2}$ (BCP) has been recently identified as a new solar absorber with promising optoelectronic properties. This work demonstrates that, despite having a low precursor purity (98.90\% to 99.95\%), synthesized BCP samples exhibit a promising photoconductive carrier lifetime up to 300 ns, an implied open-circuit voltage exceeding 1 V, {\color{black} and photoluminescence quantum yield in the order of $0.2\%$, comparable to a high-purity single-crystalline GaAs wafer. To better understand the underlying mechanisms of BCP's promising properties, its tolerance to intrinsic defects and extrinsic impurities is investigated using first-principles defect modeling and compared with that of the} well-studied GaAs. The results show that the nonradiative recombination rates induced by dominant deep-level intrinsic antisite defects are lower in BCP than in GaAs under typical growth conditions. Further exploration of the impact of transition metal impurities in the raw materials used to make BCP and impurities introduced during its synthesis shows that most of these do not form deep-level nonradiative recombination centers. As an impurity-tolerant counterpart of GaAs, BCP demonstrates great potential to improve the cost-to-performance ratio of photovoltaics.
\end{abstract}
% \secsep

{\color{black} The photovoltaic (PV) field is currently dominated by crystalline silicon, which is approaching its single-junction theoretical efficiency limit \cite{Green2009TheEvolution}. Meanwhile, thin-film technologies such as CdTe \cite{Burst2016CdTeBarrier}, Cu(In,Ga)Se$_2$ (CIGS) \cite{Jackson2011New20}, and amorphous silicon (a-Si) \cite{Rech1999PotentialCells} have been developed to reduce material usage and manufacturing costs. More recently, emerging absorbers like halide perovskites \cite{Kojima2009OrganometalCells}, kesterites ($\mathrm{Cu_2ZnSnS_4}$, CZTS) \cite{Wang2014DeviceEfficiency}, and selenium \cite{Todorov2017UltrathinMaterial} have garnered significant attention for their potential to enable high-efficiency, low-cost, or flexible devices. Amid this diverse material landscape, Zintl phosphide $\mathrm{BaCd_2P_2}$ (BCP)} has recently been proposed as a promising new solar absorber. It was identified as a stable, potentially low-cost semiconductor with a long theoretical carrier lifetime, high absorption coefficient, and reasonable carrier mobilities (comparable to CdTe) through a large-scale computational screening of 40,000 inorganic materials \cite{Yuan2024DiscoveryAbsorber}. The long carrier lifetime in BCP has been attributed to it having few low-formation energy, deep intrinsic defects. In addition, BCP has a direct band gap of $1.46\ \mathrm{eV}$ and a high computed optical absorption coefficient ($>$$10^4\ \mathrm{cm}^{-1}$) in the visible-light range. Experimentally, phase-pure BCP powder samples were synthesized and found to exhibit strong band-to-band photoluminescence (PL) emissions and a carrier lifetime of up to $30\ \mathrm{ns}$ at room temperature. {\color{black} While the optoelectronic potential of BCP is significant, the toxicity of Cd requires careful consideration for future deployment; we discuss this in more detail in our preceding work \cite{Yuan2024DiscoveryAbsorber}. Compared to some emerging hybrid perovskites and Cd-containing PV technologies (such as CdTe), which have been commercially deployed and life-cycle assessments have concluded that risks during normal use can be low when appropriate controls are in place \cite{Fthenakis2003CdTeRisks, Fthenakis2007CdTeComparisons}, BCP exhibits much stronger chemical and thermal stability. BCP is stable in air, moisture, and alkaline solutions, and has good thermal stability, with no photoluminescence degradation after thermal cycling at 368 K in air and no dissociation observed up to 1000 K \cite{Yuan2024DiscoveryAbsorber}. BCP's stability may reduce degradation-driven heavy metal release compared to other emerging absorber materials that are less chemically robust. We also note that we have investigated Cd-free analogues, particularly, CaZn$_2$P$_2$, for which we have developed crystals and thin films \cite{Quadir2024LowTemperatureAbsorbers}.} The discovery of BCP unveils a class of $\mathrm{AM_2P_2}$ (A=Ca, Sr, Ba, M=Zn, Cd, Mg) Zintl-phosphide absorbers with band gaps in the range of $\sim$$1.5 - 2.0\ \mathrm{eV}$ \cite{Quadir2024LowTemperatureAbsorbers, Hautzinger2025SynthesisApplications, Kauzlarich2023ZintlMaterials,Pike2025AStructure,Zayed2025StudyEnergy, Pike2026AlloyingCompounds}.

In this study, we evaluate the optoelectronic quality of BCP and compare it to that of GaAs, a well-studied standard for high-efficiency inorganic solar absorbers. Both BCP and GaAs have a direct band gap that is nearly optimal for single-junction photovoltaics (PV) \cite{Yuan2024DiscoveryAbsorber,Blakemore1982SemiconductingArsenide}. We synthesized BCP powder using solid-state reactions (termed ``BCP-Powder") and grew small BCP crystals with a volume of $\sim$$1\ \mathrm{mm^3}$ using Sn flux (``BCP-Crystal"). The detailed synthesis and phase confirmation procedures can be found in \hyperref[sec:s1]{Section S1} of the supporting information (SI). The purity of the elements used in the synthesis of the BCP samples ranges from 98.90\% to 99.95\% based on the Certificate of Analysis from Alfa Aersa \cite{ThermoFisherScientificInc.BariumAnalysis}, barely above the metallurgical grade and much lower than the purity of single-crystalline semiconductor wafers. We first evaluate the rate of surface recombination losses for native, unpassivated surfaces of BCP by comparing the PL intensity of BCP-Powder with a powder of GaAs prepared by grinding a prime-grade single-crystal wafer (``GaAs-GW") having an Si doping concentration of $1.6 \times10^{17} \ \mathrm{cm}^{-3}$ and resistivity of $2.2 \times 10^{-2}\ \Omega\ \mathrm{cm}$. Subsequently, we compare the implied open-circuit voltage ($V_\mathrm{OC}$) — which is derived from the quasi-Fermi level splitting \cite{Unold2022AcceleratingMaterials, Caprioglio2019OnCells,Adeleye2021LifetimeAbsorbers}  —  and carrier lifetime of BCP-Crystal with those of a pristine (unground) prime-grade wafer of GaAs (``GaAs-Wafer"). Finally, we perform first‐principles point defect calculations to investigate the impact of both intrinsic and extrinsic point defects in BCP. We compare the Shockley-Read-Hall (SRH) nonradiative recombination rate resulting from the major deep-level defects in BCP and GaAs, then assess the potential impact of extrinsic point defects in BCP arising from impurities from the raw materials used for its synthesis.

\begin{figure}[!t]
    \centering
    \includegraphics[width=0.9\linewidth]{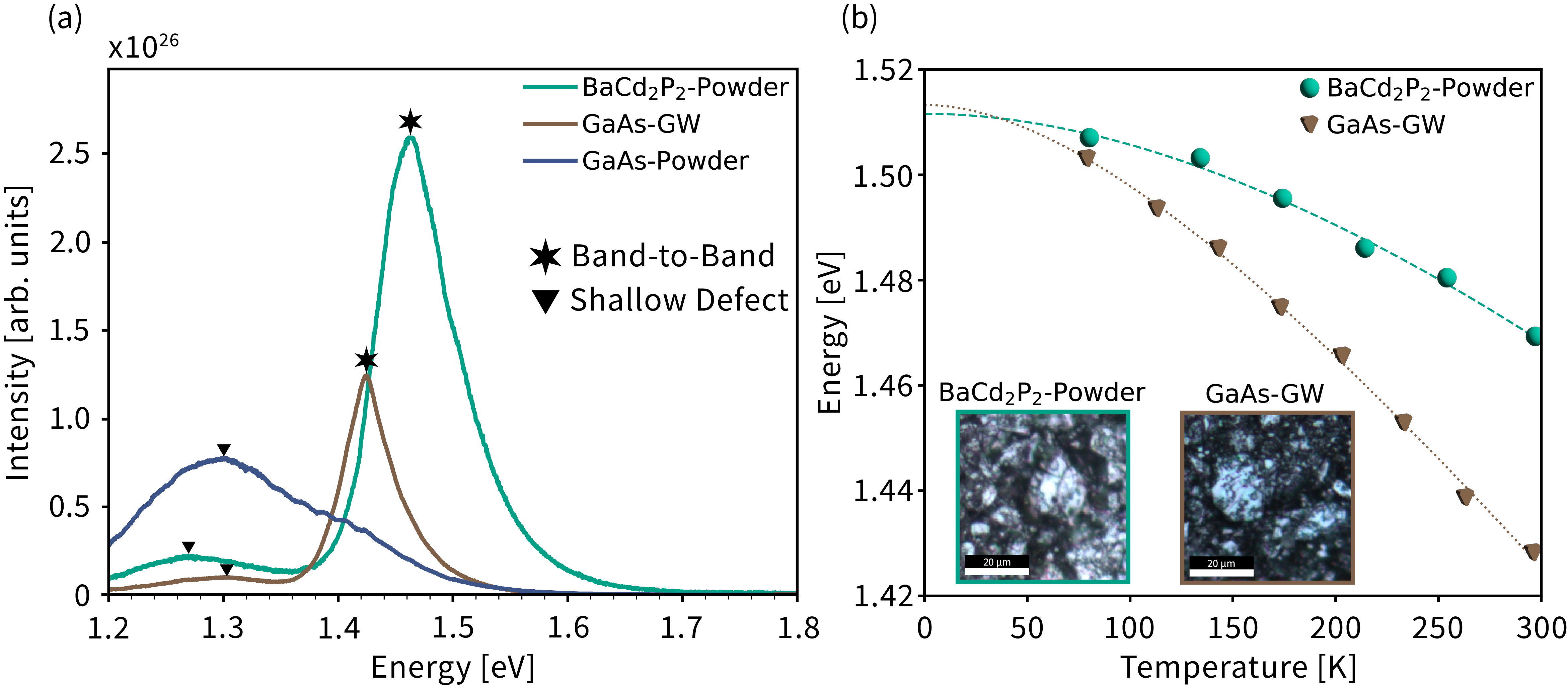}
    \caption{\label{fig:1}(a) PL spectra at $298\ \mathrm{K}$ for BCP-Powder, GaAs-Powder synthesized using solid state reactions, and GaAs-GW — a commercial 2-inch GaAs (001) single-crystalline prime wafer that has been ground into a powder. All PL spectra were collected using a $532\ \mathrm{nm}$ laser with an excitation power of $30\ \mathrm{\mu W}$ and a focal spot diameter of 3 $\mu$m incident on the surface of the samples. The broad sub-bandgap peaks near $1.3\ \mathrm{eV}$ arise from shallow defect emissions \cite{Klingshirn2007SemiconductorOptics, Bhattacharya2011ComprehensiveTechnology}. (b) Temperature-dependent PL tracking the shift in the band-to-band emission peak position of BCP-Powder and GaAs-GW; the shift is fit using Varshni’s equation \cite{Varshni1967TemperatureSemiconductors} to determine the 0 K band gap as discussed in \hyperref[sec:s2]{SI-Section 2}. The inserts in (b) are the optical microscope images of BCP-Power and GaAs-GW (the scale bar represents $20\ \mathrm{\mu m}$).}
\end{figure}

\textbf{Photoluminescence.} We begin by probing the surface recombination sensitivity of BCP and GaAs in terms of their native, unpassivated surfaces. Surface recombination reduces diffusion length and $V_\mathrm{OC}$, greatly lowering the performance of minority-carrier devices like solar cells \cite{Nelson1980ReductionTreatment}. To see the impact of native surface recombination, we look at the room-temperature PL spectra of BCP-Powder and GaAs-GW. The high surface-to-volume ratio of the unpassivated powder samples makes their PL intensity sensitive to surface recombination losses, thus making PL intensity analysis a good metric for comparison. As illustrated in \hyperref[fig:1]{Figure 1a}, BCP-Powder has a more intense band-to-band emission peak at $1.46\ \mathrm{eV}$ compared to GaAs-GW whose corresponding peak is at $1.42\ \mathrm{eV}$ (here, the same excitation laser power of $30\ \mu \mathrm{W} $ is used for both samples). The contrast in emission intensities is impressive when considering that BCP-Powder was prepared using low-purity precursors with transition metal impurities on the order of tens of ppm with a solid state synthesis technique prone to other impurities and additional defects (e.g., carbon, oxygen, silicon from the carbonized silica ampule \cite{Chattopadhyay1998CharacterizationTransport, Fang1989AnnealingGaAs, Thorne1992EffectNbSe3, Egarievwe2016CarbonDetectors, Shetty1995InfluenceSolidification}), while GaAs-GW was sourced from a commercial prime-grade single-crystal wafer. As confirmation, we also synthesized BCP-Powder from several batches of (low-purity) Ba and Cd from different vendors, and found that it reproduces similar PL performance. 

To eliminate the difference in synthesis quality between BCP-Powder and GaAs-GW as a factor in the comparison, we prepared a GaAs powder using solid-state reactions similar to those used for BCP-Powder (“GaAs-Powder”). In this case, the purity levels of the Ga and As sources are 99.999\% and 99.9999\%, respectively, both much higher than that of the raw materials for BCP. Interestingly, we find that GaAs-Powder shows little room-temperature PL peak corresponding to band-to-band emission (\hyperref[fig:1]{Figure 1a}); its PL spectrum is dominated by a broad shallow defect emission peak at $1.3\ \mathrm{eV}$. This comparison suggests that BCP is less susceptible to surface recombination and more tolerant to impurities introduced during synthesis than GaAs. 

We also compare the PL intensity of BCP-Crystal and pristine GaAs-Wafer (note that the raw material quality and Sn-flux preparation of BCP-Crystal may introduce many more unwanted impurities compared to the prime grade GaAs-Wafer). \hyperref[fig:s1]{Figure S1} shows that GaAs-Wafer has brighter PL emission than BCP-Crystal, supporting the observation that increased surface recombination resulting from the higher surface area of the powders affects GaAs more than BCP. The ability of BCP to yield bright band-to-band emission despite its low-purity raw material precursors and impurity-prone synthesis highlights its potential compatibility with low-cost PV manufacturing.

\hyperref[fig:1]{Figure 1a} also shows that BCP and GaAs have similar band gaps at room temperature - note that the PL peak position overestimates the band gap by $\sim$$\mathrm{k_B T}/2$ as shown in\hyperref[fig:s3]{Figure S3} \cite{FredSchubert2016LightDiodes}. The similarity of the two materials becomes more evident when examining their $0\ \mathrm{K}$ band gaps extracted by fitting the temperature dependence of the PL peak positions to Varshni’s equation \cite{Varshni1967TemperatureSemiconductors} (details are in \hyperref[sec:s2]{SI-Section 2}). As seen in \hyperref[fig:1]{Figure 1b}, the extrapolated $0\ \mathrm{K}$ band gap is $\approx$$1.51\ \mathrm{eV}$ for both. Our GaAs result agrees well with literature values, validating the accuracy of our temperature‐dependent PL measurements \cite{Levinshtein1996HandbookParameters, Blakemore1982SemiconductingArsenide}.

\textbf{Implied Open-circuit Voltage.} $V_\mathrm{OC}$ is an important metric of solar cell device performance, directly reflecting the maximum attainable voltage under illumination, and it can be limited by nonradiative carrier recombination \cite{Leggett2012TheNelson, Siebentritt2022PhotoluminescenceAbsorbers, Siebentritt2021HowCells}. Here, we compare the implied $V_\mathrm{OC}$, {\color{black} i.e., the $V_\mathrm{OC}$ in the ideal case, ignoring factors such as series resistance and contact losses, for} BCP and GaAs. Implied $V_\mathrm{OC}$ is directly related to the electron and hole quasi-Fermi level splitting ($\Delta\mu$) upon illumination according to the relation $V_\mathrm{OC}=\Delta\mu / q$, where $q$ is the elementary charge \cite{Unold2022AcceleratingMaterials, Caprioglio2019OnCells,Adeleye2021LifetimeAbsorbers}. {\color{black} Note that the implied $V_\mathrm{OC}$ does not make any considerations to resistive transport losses, and thus represents the maximum achievable $V_\mathrm{OC}$.} $\Delta\mu$ can be obtained by fitting the PL peak corresponding to the band-to-band emission to the spontaneous emission equation modified for nonequilibrium conditions \cite{PWurfel1982TheRadiation, Lasher1964SpontaneousSemiconductors, Katahara2014Quasi-FermiPhotoluminescence, Fadaly2020Direct-bandgapAlloys, Unold2016PhotoluminescenceCells, vanRoosbroeck1954Photon-RadiativeGermanium}:

\begin{figure}[!t]
    \centering
    \includegraphics[width=0.9\linewidth]{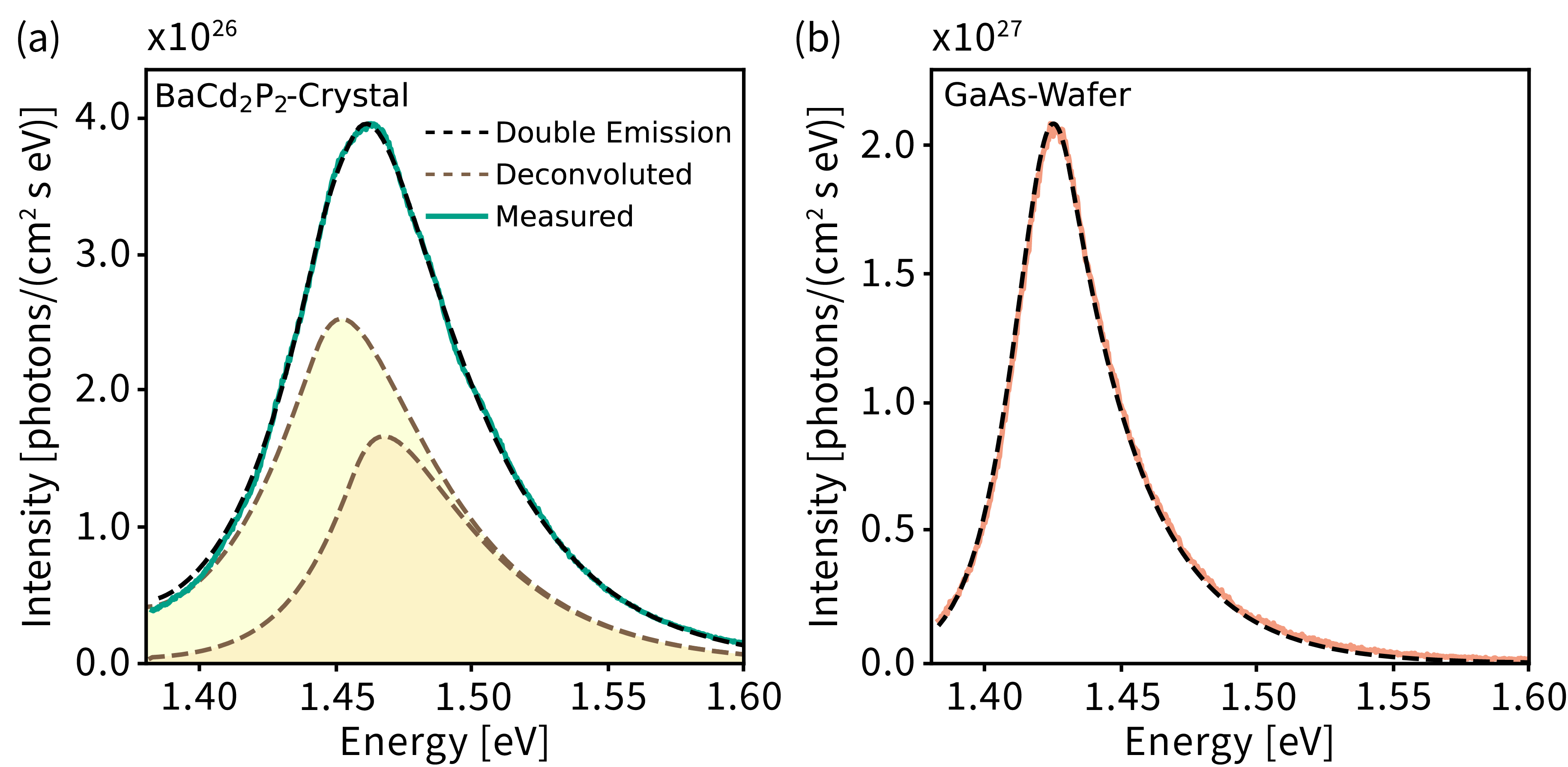}
    \caption{\label{fig:2}Fitting the 298 K PL spectra of (a) BCP-Crystal and (b) GaAs-Wafer to the spontaneous emission equation for nonequilibrium conditions (\hyperref[eq:1]{Equation 1}) using an incident excitation power density of $354 \ \mathrm{W/cm^2}$ and $159\ \mathrm{W/cm^2}$ respectively. The black dashed lines in both (a) and (b) represent the overall spectra obtained from the fitting and agree well with the measured PL. The GaAs-Wafer peak is a result of a single direct gap transition, while that of BCP-Crystal is the result of two band-to-band transitions. The brown dashed lines in (a) illustrate the deconvolution of the band-to-band transitions.}
\end{figure}

\begin{equation}
I_{PL} = \frac{2 \pi}{c^{2}h^{3}}\frac{E^{2}}{\exp{\left( \frac{E - \mathrm{\Delta}\mu}{k_BT} \right) - 1}} a(E)
\label{eq:1}
\end{equation}
{\color{black} where $I_{\rm PL}$ is the absolute PL intensity in units of $\frac{\mathrm{Photons}}{\mathrm{cm}^2 \ \mathrm{s} \ \mathrm{eV}}$, $a(E)$ is the absorptivity at photon energy $E$ (taking into account the effect of quasi-Fermi level separation), $c$ is the speed of light, and $h$ is Planck’s constant. Sub-band gap absorption from the Urbach tail is included in $a(E)$ as explained in \hyperref[sec:s3]{SI-Section 3.1-3.3}. The instrument was calibrated using the Raman intensity of Si to obtain absolute PL intensity (details are provided in \hyperref[sec:s3]{SI-Section 3.4}). The PL fitting is restricted to band-to-band transitions, ensuring that defect-related sub-band-gap emissions near $1.3\ \mathrm{eV}$ are excluded. Furthermore, as shown in \hyperref[fig:s3]{Figure S3}, the fitting of $\Delta \mu$ is mostly determined by the exponential decay tail of the Bose-Einstein term on the higher energy side of the PL spectrum, which is far from the Urbach tail. 

When performing the PL fitting, we reduced the number of free parameters by using the steady-state excess carrier density ($\Delta n$) as a fitting parameter rather than determining the quasi- hole and electron Fermi levels. We then derive $\Delta \mu$ from $\Delta n$ and the first-principles computed carrier effective masses (a detailed discussion of how the fitting was performed is provided in \hyperref[sec:s3]{SI-Section 3.2}). Note that the excitation power density sets the photogeneration rate, but not $\Delta n$; the latter also depends on the effective carrier lifetime, which is unknown a priori. Thus $\Delta n$ is treated as a fitting parameter. For each material, we perform a global fit to multiple PL spectra collected over a range of excitation power densities rather than fitting each PL spectrum independently. In this global fit, the material-specific quantities such as the Urbach energies ($E_U$) and band gaps ($E_g$) are constrained to be shared across all spectra, while only the power-dependent quantity $\Delta n$ is allowed to vary from spectrum to spectrum. As a robustness check, in addition to the full model presented in \hyperref[sec:s3]{SI-Section 3.1}, we also fit the PL using a simplified absorptivity model with fewer fitting parameters (as discussed in \hyperref[sec:s3]{SI-Section 3.3}; Case I for BCP and Case II for GaAs). The extracted $\Delta\mu$ values (and therefore implied $V_\mathrm{OC}$) are nearly unchanged across the two models, indicating that the main conclusions are not sensitive to the additional flexibility of the full absorptivity model. $E_U$, $E_g$, and other fitted parameters from \hyperref[eq:1]{Equation 1} are presented in \hyperref[sec:s3]{SI-Section 3.3}}.

For a single band-to-band transition, the PL full width at half-maximum (FWHM) is $\sim$$1.8\ k_\mathrm{B}T$ when assuming a parabolic band model \cite{FredSchubert2016LightDiodes}. The $1.42\ \mathrm{eV}$ GaAs peak follows this trend, but the $1.46\ \mathrm{eV}$ peak for the BCP-Crystal has a FWHM of $1.5 \times 1.8\ k_\mathrm{B}T$ (\hyperref[table:s1]{Table S1}). The BCP-Crystal peak is broader than expected for a single emission, suggesting it is composed of more than one band-to-band emission. We find that fitting the PL spectrum of BCP-Crystal using \hyperref[eq:1]{Equation 1} to two emissions reproduces the measured BCP-Crystal PL spectrum better than a single emission. \hyperref[fig:2]{Figure 2a} shows the PL spectrum of BCP-Crystal fit to \hyperref[eq:1]{Equation 1} assuming two band-to-band emissions, and reveals that the first emission has an energy of $1.44\ \mathrm{eV}$ and the second $1.46\ \mathrm{eV}$. These emissions likely originate from the two closely situated conduction band valleys at the $\Gamma$ point in the electronic band structure of BCP (see \hyperref[fig:s6]{Figure S8}). Both emissions experience the same thermal broadening and share one $\Delta\mu$ to satisfy carrier quasi-equilibrium; we obtain the $\Delta\mu$ by iteratively solving the charge-balance equations as discussed in \hyperref[sec:s3]{SI-Section 3}. Performing the \hyperref[eq:1]{Equation 1} fit for GaAs-Wafer gives a single emission with a bandgap of $1.42\ \mathrm{eV}$, consistent with expectations (\hyperref[fig:2]{Figure 2b}). We then use $\Delta\mu$ to determine the implied $V_\mathrm{OC}$ for the two materials.

\begin{figure}[!t]
    \centering
    \includegraphics[width=0.64\linewidth]{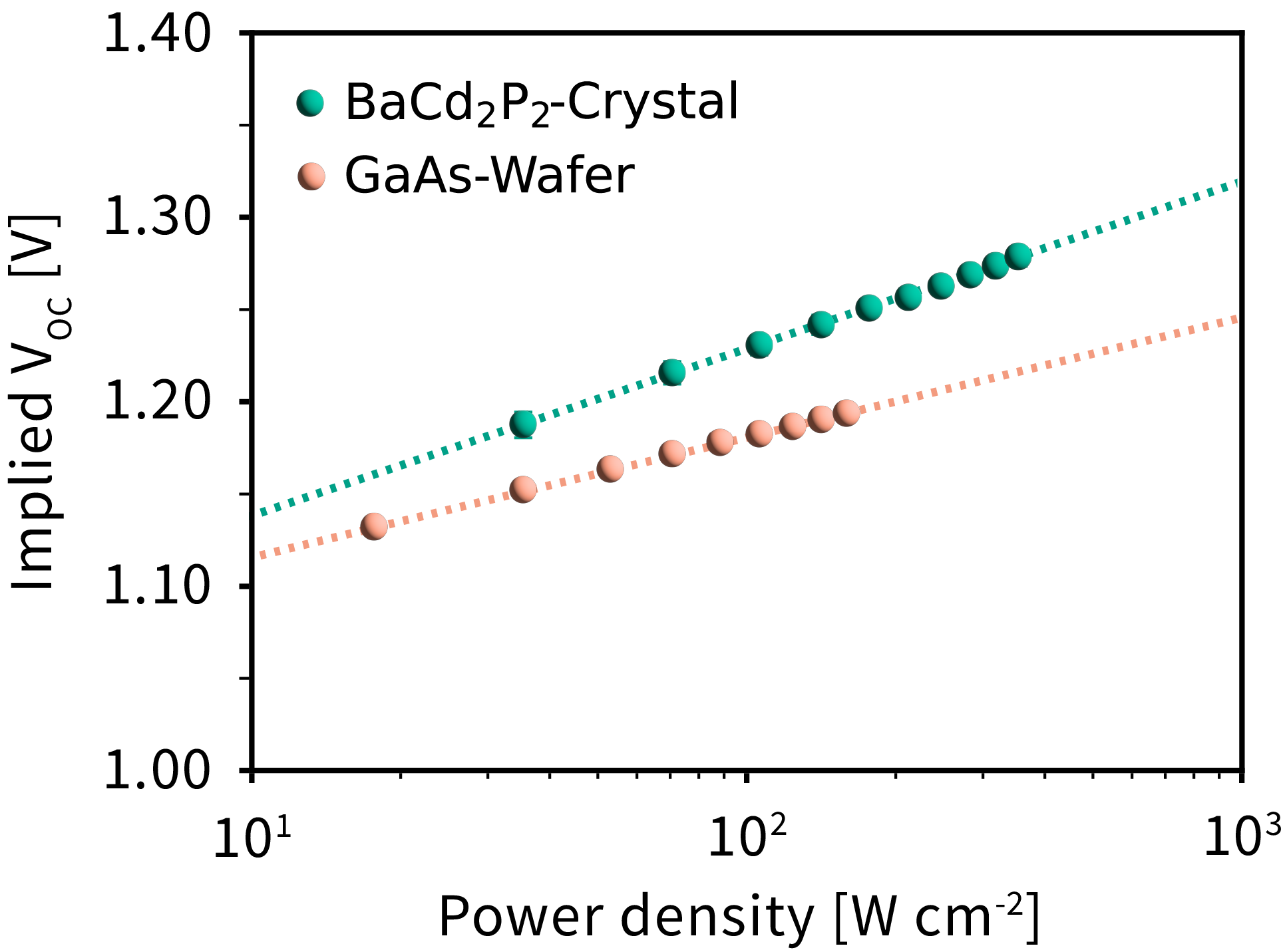}
    \caption{Implied V$_\mathrm{{OC}}$ vs incident optical power density of BCP-Crystal and GaAs-Wafer. The dashed lines correspond to linear fits. Note that the error bars are smaller than the size of the symbols for both materials. The measurements were performed at 298 K.}
    \label{fig:3}
\end{figure}

We obtained the evolution of the implied $V_\mathrm{OC}$ as a function of incident optical power density of a 532 nm wavelength laser by collecting the PL spectra under varying laser powers and fitting them to \hyperref[eq:1]{Equation 1} (\hyperref[fig:3]{Figure 3}). We note that, while 532 nm is also the peak emission wavelength of the solar spectrum, one should not directly compare the PL power density with AM1.5 solar power density because the photon fluxes are different. At the same power density, the photon flux at 532 nm excitation wavelength is 1.34x greater than that of the AM1.5 solar spectrum at photon energies above $1.42\ \mathrm{eV}$, i.e., the bandgap of GaAs.  We find that BCP-Crystal has an implied $V_\mathrm{OC}$ consistently higher than that of the GaAs-Wafer in the tested power density range, and this observation is further supported by \hyperref[fig:s1]{Figure S1}, which shows that the high-energy Bose-Einstein decay tail of BCP-Crystal extends further than that of GaAs-Wafer. 

{\color{black}The higher implied $V_\mathrm{OC}$ of BCP-Crystal compared to GaAs-Wafer is partly due to BCP having a larger band gap at the measurement temperature than GaAs; $\sim$$40\ \mathrm{mV}$ of the $V_\mathrm{OC}$ difference is due to the band gap difference. In the ideal case, $V_\mathrm{OC}$ would decrease by $\ln(10)\ \mathrm{k_B T}/q$ per decade decrease in optical power density. The slope of the $V_\mathrm{OC}$ vs. power density plot of BCP-Crystal is $90\ \mathrm{mV/decade}$, and this corresponds to an ideality factor of 1.43 at $315\ \mathrm{K}$ (the apparent surface temperature obtained from BCP-Crystal's PL fitting)}. By comparison, the implied $V_\mathrm{OC}$ of single crystalline GaAs-Wafer increases by $64\ \mathrm{mV/decade}$, corresponding to an ideality factor of 1.08 at its PL fitting extracted temperature of 300 K. The higher ideality factor in BCP-Crystal suggests a greater degree of trap-assisted recombination in BCP-Crystal, which is consistent with the much higher impurity concentration of the BCP-Crystal sample compared to GaAs-Wafer (trap-assisted recombination reduces the steady-state carrier population in the bands, and correspondingly $V_\mathrm{OC}$). The larger slope of BCP-Crystal also implies that extrapolating the $V_\mathrm{OC}$ vs. power density plot to 1 Sun illumination would lead to similar $V_\mathrm{OC}$ values for BCP and GaAs. Overall, it is very impressive that a low-purity (nearly metallurgical grade) BCP-Crystal achieved a similar implied $V_\mathrm{OC}$ as the commercial single crystalline GaAs-Wafer. BCP's tolerance to impurities could potentially reduce the fabrication costs while maintaining a large $V_\mathrm{OC}$.

% The standard condition for reporting solar cell $V_\mathrm{OC}$ is $\mathrm{AM1.5}$, which corresponds to a power density of $\sim 0.1\ \mathrm{W}/\mathrm{cm^2}$ \cite{Saga2010AdvancesProduction}. Extrapolating the values in \hyperref[fig:3]{Figure 3} to $\mathrm{AM1.5}$ gives an implied $V_\mathrm{OC}$ that is $1.1\ \mathrm{V}$ for BCP-Crystal and $0.95\ \mathrm{V}$ for GaAs-Wafer. Our results for GaAs are close to previous literature reports of $\sim1\ \mathrm{V}$, especially when considering the apparent temperature of the wafer used here is $320\ \mathrm{K}$ and no lifetime optimizations or photon recycling engineering were preformed here   \cite{Steiner2013OpticalCells,Tobin1990AssessmentApplications, Sodabanlu2018ExtremelyApplication, Lang2018OptimizationUm/h}.

\begin{figure}[!t]
    \centering
    \includegraphics[width=0.9\linewidth]{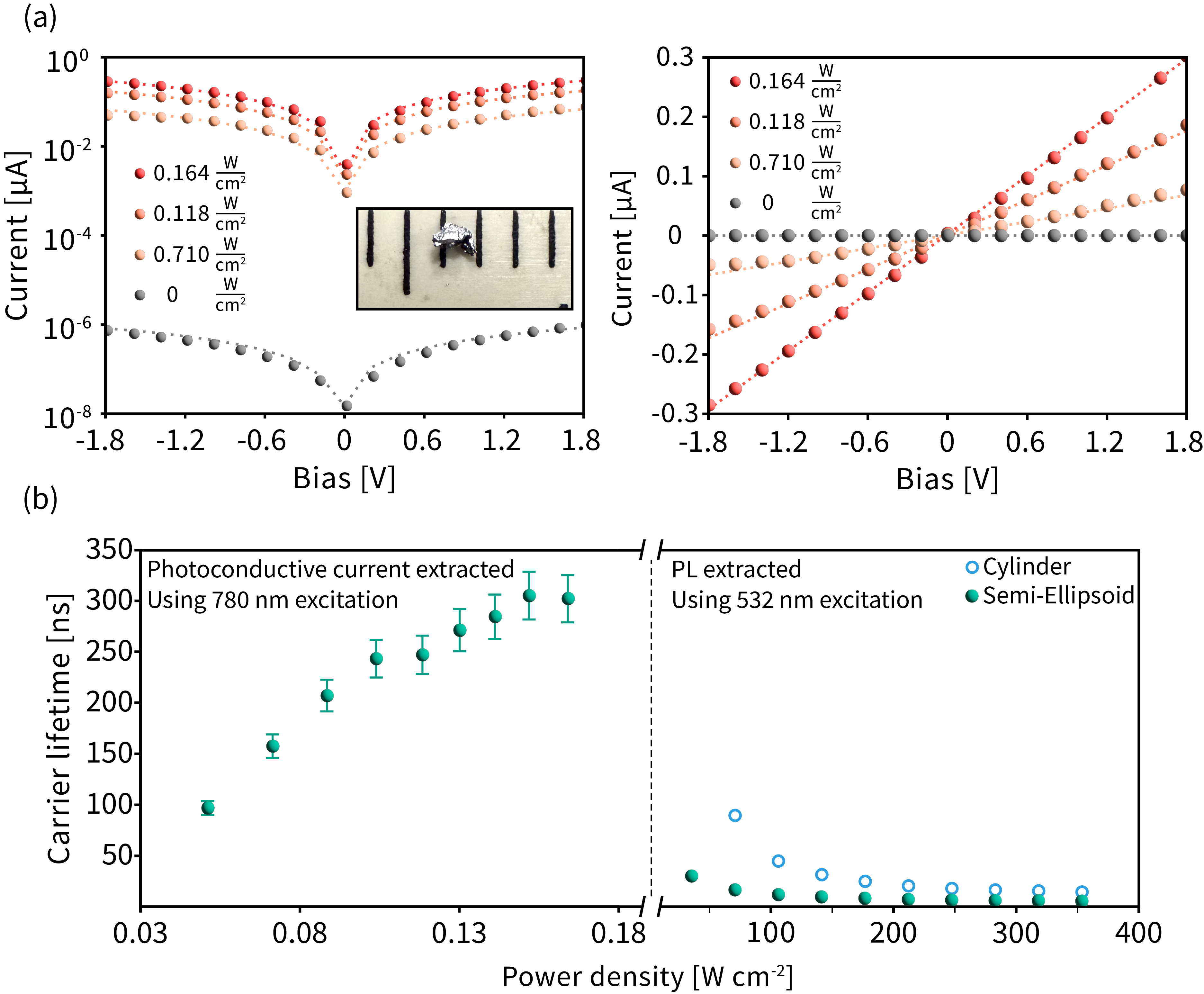}
    \caption{\label{fig:4}(a) Magnitude of photoconductive current for BCP-Crystal at three illumination levels of a $780\ \mathrm{nm}$ laser diode compared with the dark current ($0\ \mathrm{W/cm^2}$). The dashed lines correspond to linear fits. The minimum of the dark current at $0\ \mathrm{V}$ bias was  $0.1\ \mathrm{pA}$, which is the noise floor of the system. Accordingly, the current value at $0\ \mathrm{V}$ has been set to $0\ \mathrm{A}$. The inset shows a BCP-Crystal placed on a ruler with $1\ \mathrm{mm}$ tick spacing. (b) The effective carrier lifetime of BCP-Crystal from $0.05$ to $0.164\ \mathrm{W/cm^2}$ extracted from photoconductive current measurements (left) and from $35$ to $354\ \mathrm{W/cm^2}$ extracted from the PL fit of \hyperref[eq:2]{Equation 2} using a 532 nm laser (right). The error bars are smaller than the size of the symbols for the PL extracted carrier lifetime (right). All the measurements were carried out at $298\  \mathrm{K}$.}
\end{figure}

\textbf{Photoconductive Lifetime.} We find that BCP-Crystal has impressive photoresponse, where, upon illumination using a $780\ \mathrm{nm}$ laser diode at a power density of $0.164\ \mathrm{W/cm^2}$, the photoconductive current is around $\mathrm{10^5}$ times greater than the dark current (\hyperref[fig:4]{Figure 4a}). We achieved Ohmic contact with the BCP-Crystal using indium pads (\hyperref[fig:s5]{Figure S5}) and confirmed there was negligible contact resistance (\hyperref[fig:s6]{Figure S6}) as outlined in \hyperref[sec:s4]{SI-Section 4}. Such small contact resistance means that heating losses at the contacts are virtually eliminated, and, as a result, future device design and integration will be greatly simplified. 

Here, we estimate the carrier lifetime ($\tau$) of BCP-Crystal and GaAs-Wafer using their photoconductive and dark currents. For a piece of semiconductor with electrodes fully contacting both sidewalls (as shown in \hyperref[fig:s5]{Figure S5}), $\tau$ can be obtained using \cite{Sze2006PhysicsDevices}:
\begin{equation}
\tau=\frac{\Delta n}{G_e}
\label{eq:2}
\end{equation}
where $G_e$ is the generation rate of photo-excited electrons and $\Delta n$ is the injected carrier density. We assume that carrier mobilities do not change significantly {\color{black}at low injection levels, far from degeneracy, which is reasonable given that the dominant impurity and photon scattering mechanisms remain unchanged, whereas} free-carrier scattering is negligible for non-degenerate injection. Therefore, the mobilities are considered the same in dark and under $0.164\ \mathrm{W/cm^2}$ illumination. Since there is also negligible contact resistance in the case of BCP-Crystal (and similarly the contact resistance is 5 orders lower than the sample resistance under illumination in GaAs-Wafer as shown in \hyperref[sec:s4]{SI-Section 4.2}), the lower limit of $\Delta n$ can be determined by:
\begin{equation}
\Delta n =n_\mathrm{dark}\ \left(\frac{R_\mathrm{dark}}{R_\mathrm{light}}-1\right),
\label{eq:3}
\end{equation}
where $R_\mathrm{dark}$ and $R_\mathrm{light}$ represent the resistance in dark and under light illumination, respectively, $n_\mathrm{dark}$ is the carrier concentration in dark. We assume $n_\mathrm{dark}=n_i$ in the following analyses, where $n_i$ is the intrinsic carrier density. Since any residual doping would increase the carrier concentration beyond the intrinsic carrier density, assuming $n_\mathrm{dark}=n_i$ can only underestimate $n_\mathrm{dark}$, and as a result, underestimate $\Delta n$ and $\tau$ according to \hyperref[eq:3]{Equation 3}. As discussed in \hyperref[sec:s4]{SI-Section 4}, we use $n_i \approx 2 \times 10^7 \ \mathrm{cm^{-3}}$ for BCP-Crystal, obtained using PL fitting data and by solving the charge balance condition while considering point defects. Finally, $G_e$ is given by \cite{Sze2006PhysicsDevices}:
\begin{equation}
G_e =\frac{\eta \ (\frac{P_{\mathrm{opt}}}{hv})}{V}
\label{eq:4}
\end{equation}
where $\eta$ is the external quantum efficiency of electron-hole pair generation, $P_{\mathrm{opt}}$ is the incident optical power, $hv$ is the photon energy, and $V$ is the sample volume under illumination. We can then use these $G_e$ and $\Delta n$ values in \hyperref[eq:2]{Equation 2} to determined $\tau$. In practice, we use the equivalent relation given by \hyperref[eq:s47]{Equation S47}, which has the advantage of having the sample width and thickness cancel out (also discussed in \cite{Sze2006PhysicsDevices}). Doing so leaves imprecision in measuring the length of the crystal in the direction of the current flow, measured to be $1.25\ (\pm\ 0.05)\ \mathrm{mm}$, as the primary source of uncertainty in determining $\tau$.

We obtain $\tau$ of BCP-Crystal at several laser power densities. \hyperref[fig:4]{Figure 4b} shows that $\tau$ initially increases with increasing power density before saturating around $0.15\ \mathrm{W/cm^2}$ at a value of $\tau \geq 300 \ \mathrm{ns}$ — the initial increase is likely due to the filling of trap states. This value is a lower limit for $\tau$, because we set $\eta=1$, and doing so overestimates $G_e$, and therefore, underestimates $\tau$. Note that even this lower limit is an order of magnitude higher than the value of $30\ \mathrm{ns}$ previously measured for BCP-Powder using time-resolved microwave conductivity \cite{Yuan2024DiscoveryAbsorber}. For photon flux above the band gap equivalent to 1 Sun in AM1.5 spectrum, the corresponding carrier lifetime is approximately 100 ns for $0.50 \ \mathrm{W/cm^2}$ at 780 nm excitation. By comparison, for GaAs-Wafer, we found $\tau$ to be $\approx5\ \mathrm{ns}$ at $0.164 \ \mathrm{W/cm^2}$; this value of $\tau$ is in line with literature values for heavily Si-doped GaAs \cite{P.A.Folkes2012MinorityHeterostructures, Lush1992ADeposition, Tarricone1986HoleBarriers, Hwang1971DopingGaAs}. {\color{black} It is important to note that part of the reason for the lower lifetime of GaAs is its very large radiative recombination rate, which is higher than that of other III-V semiconductors \cite{Niemeyer2019MeasurementGaAs}. For instance, \cite{Strauss1993AugerGaAs} report a radiative recombination rate of $B=(1.7\pm0.2)\times10^{-10}\ \mathrm{cm^3/s}$, which would lead to a lifetime of $36.8\pm 4.3\ \mathrm{ns}$ for our $1.6 \times 10^{17}\ \mathrm{cm^{-3}}$ doped GaAs. Similarly we obtain $18.9\pm 3.4\ \mathrm{ns}$ when using the rate $(B=3.3\pm0.6)\times10^{-10}\ \mathrm{cm^3/s}$ reported by \cite{Niemeyer2019MeasurementGaAs}. These values suggest that the fast radiative recombination contributed significantly to the measured short carrier lifetime in GaAs.}

We then determine the sum of the electron mobility ($\mu_n$) and the hole mobility ($\mu_p$) for BCP-Crystal using the resistance at any pair of $P_\mathrm{{opt}}$ levels $P_1$ and $P_2$ as follows:
\begin{equation}
\mu_n + \mu_p = \frac{1-\frac{P_1}{P_2}}{n_i\frac{qA}{L}[(R_\mathrm{dark}^\mathrm{meas}-R_\mathrm{light_2}^\mathrm{meas})-\frac{P_1}{P_2}(R_\mathrm{dark}^\mathrm{meas}-R_\mathrm{light_1}^\mathrm{meas})]}
\label{eq:5}
\end{equation}
where $R_\mathrm{light_1}^\mathrm{meas}$ and $R_\mathrm{light_2}^\mathrm{meas}$ are the measured resistances (i.e., considering contact resistance) at $P_1$ and $P_2$ respectively, $q$ is the elementary charge, $A$ is the cross-sectional area, and $L$ is the sample length (details and derivation in \hyperref[sec:s4]{SI-Section 4}). For BCP-Crystal, we find $\mu_n + \mu_p  \approx 5\ \mathrm{cm^2/Vs}$, and this value is in line with the TRMC data reported in \cite{Yuan2024DiscoveryAbsorber}. The mobility of BCP-Crystal is relatively lower than the theoretical predictions in \cite{Yuan2024DiscoveryAbsorber} and that of GaAs-Wafer, whose $\mu_n$ was found to be $1740\ \mathrm{cm^2/Vs}$ using Hall effect measurements. The lower mobility of BCP-Crystal is likely a result of scattering of carriers by the relatively high concentration of impurities from the raw materials and the fabrication process, rather than intrinsic behavior. Using the measured $\mu_n + \mu_p$ and the estimated $\tau$ (300 ns at $0.15\ \mathrm{W/cm^2}$), we determine the ambipolar diffusion length $L_\mathrm{D}=\sqrt{\frac{\mu k_\mathrm{B}T}{e}\tau}$ to be 2 $\mu$m, which is high enough to enable efficient carrier collection for thin film photovoltaic cells \cite{Akel2023RelevanceCells}.

Next, we determine $\tau$ at the PL excitation power, which is 2-3 orders higher than the photoconductivity measurements, using $\Delta n$ obtained from the fit of \hyperref[eq:1]{Equation 1} (where $\Delta n$ is in the range of $10^{16} - 10^{17} \mathrm{cm}^{-3}$) and by calculating $G_e$. Assuming negligible photon recycling, $G_e$ can be determined using \hyperref[eq:4]{Equation 4}; in this case, $P_\mathrm{opt}$ is the laser optical power. We use two models to determine $V$, where we treat the total interaction volume as (1) a cylinder, and (2) a semi-oblate spheroid - a semi-ellipsoid with two equal axes. If treating it as a cylinder, the radius is the sum of the laser spot radius ($r_\mathrm{spot} = 1.5\  \mathrm{\mu m}$) and carrier diffusion length ($d$), while the height is the sum of $d$ and the penetration at the excitation laser wavelength ($1 / \alpha_\mathrm{532\ nm}$, where $\alpha$ is the absorption coefficient obtained from the first-principles calculation in \cite{Yuan2024DiscoveryAbsorber}). In the semi-oblate spheroid model, the radius is the same as the cylinder model, but $d + 1 / \alpha_\mathrm{532\ nm}$ is now the length of the flattened axis of the spheroid. To determine $d$ we can use the relation $d = \sqrt{D_p\ \tau} = \sqrt{\mu\ \mathrm{k_B T}\ \tau / q}$. Which, for the cylinder model, gives:
\begin{equation}
V = \pi\ \left(r_\mathrm{spot}+\sqrt{\mu\ \mathrm{k_B T}\ \tau / q}\right)^2\ \left(\sqrt{\mu\ \mathrm{k_B T}\ \tau / q} + 1 / \alpha_\mathrm{532\ nm}\right)
\label{eq:6}
\end{equation}
and for the semi-oblate spheroid model gives:
\begin{equation}
V = \frac{2\pi}{3}\ \left(r_\mathrm{spot} + \sqrt{\mu\ \mathrm{k_B T}\ \tau / q}\right)^2\ \left(\sqrt{\mu\ \mathrm{k_B T}\ \tau / q} + 1 / \alpha_\mathrm{532nm}\right)
\label{eq:7}
\end{equation}
We are now at a position to use either Equation \hyperref[eq:6]{6} or \hyperref[eq:7]{7} with $\Delta n$ and combine Equations \hyperref[eq:2]{2} and \hyperref[eq:4]{4} to solve for $\tau$. The cylinder model results in a quadratic equation while the semi-oblate spheroid results in a cubic equation; the solutions for both are plotted in \hyperref[fig:4]{Figure 4b}.

This approach serves as a high power density estimate of $\tau$. We find that for BCP-Crystal, $\tau$ decreases from $30\ \mathrm{ns}$ at $35\ \mathrm{W/cm^2}$ to $14\ \mathrm{ns}$ at $354\ \mathrm{W/cm^2}$. The lower $\tau$ obtained from PL compared to the photoconductive current $\tau$ is partly due the shallower penetration of the $532\ \mathrm{nm}$ laser used for PL (vs $780\ \mathrm{nm}$ used for the photoconductive current measurements); which implies $\tau$ extracted from PL is more affected by surface recombination. The lower $\tau$ may also be partly due to the onset of Auger recombination at the high power densities used in PL \cite{Green1984LimitsProcesses}.

{\color{black} To place the measured $\tau$ values in context, we compare BCP-Crystal with representative PV absorbers in \hyperref[table:1]{Table 1}. We note that the $\tau$ values are highly measurement and sample-dependent (e.g., surface passivation, temperature), so values in \hyperref[table:1]{Table 1} are intended as representative literature benchmarks rather than directly comparable intrinsic limits.}

\begin{table}[t]
    \centering
    {\color{black} 
    
    \caption{Comparison of optoelectronic properties of BCP with established and emerging photovoltaic materials. Values represent typical high-quality performance reported in literature. We considered the MAPbI$_{3-x}$Cl$_x$ structure to represent the family of perovskites. The values of GaAs are for the GaAs doping level used in the paper ($\sim$$10^{17}\ \mathrm{cm^{-3}}$) and not considering photon recycling. We note that the measurement method and factors such as passivation have a considerable impact on the carrier lifetime, this these values are intended for order-of-magnitude comparison.\\}
    \label{table:1}
    \setlength{\tabcolsep}{0.5pt}
    \begin{tabular}{l|c|c|c|c|c|c}
        \hline
        \hline
        Material            & Sample    & Gap      & Carrier                                                          & Band Gap                                        & $V_\mathrm{OC}$ [V]                                           & ${V_\mathrm{OC}}$ \\ 
                            & Form      & Type     & Lifetime                                                         & [eV]                                            &                                                               &  Deficit [V]      \\ 
        \hline
        BCP                 & Crystal   & Direct   & 300 ns                                                           & 1.46 \cite{Yuan2024DiscoveryAbsorber}           & $\sim$1 (implied)                                             & 0.46              \\ 
        GaAs                & Crystal   & Direct   & 19 ns \cite{Hwang1971DopingGaAs, Nelson1978Minority-carrierGaAs} & 1.42 \cite{Blakemore1982SemiconductingArsenide} & 0.994 \cite{Green2024Solar64}                                 & 0.43              \\ 
        c-Si                & Crystal   & Indirect & $>$10 ms \cite{Richter2012ImprovedSilicon}                       & 1.12 \cite{Bludau1974TemperatureSilicon}        & 0.750 \cite{Taguchi201424.7Wafer}                             & 0.37              \\ 
        CdTe                & Crystal   & Direct   & 360 ns \cite{Kuciauskas2015Minority-CarrierCdTe}                 & 1.51 \cite{Fonthal2000TemperatureCdTe}          & 1.017 \cite{Burst2016CdTeBarrier}                             & 0.49              \\ 
        MAPbI$_{3-x}$Cl$_x$ & Crystal   & Direct   & 82 µs \cite{Dong2015Electron-holeCrystals}                       & 1.61 \cite{Dong2015Electron-holeCrystals}       & 0.62 \cite{Dong2015Electron-holeCrystals}                     & 0.99              \\ 
        MAPbI$_{3-x}$Cl$_x$ & Thin film & Direct   & 270 ns \cite{Stranks2013Electron-HoleAbsorber}                   & 1.60 \cite{Liu2019Open-CircuitCells}            & 1.26 \cite{Liu2019Open-CircuitCells}                          & 0.34              \\ 
        CIGS                & Thin film & Direct   & 250 ns \cite{Metzger2008LongCells}                               & 1.14 \cite{Ramanujam2017CopperReview}           & 0.71--0.74 \cite{Jackson2011New20, Ramanujam2017CopperReview} & $\sim$0.41        \\ 
        CZTS                & Thin Film & Direct   & 5.4 ns \cite{Yan2018Cu2ZnSnS4Treatment}                          & 1.51 \cite{Yan2018Cu2ZnSnS4Treatment}           & 0.730 \cite{Yan2018Cu2ZnSnS4Treatment}                        & 0.78              \\ 
        \hline
    \end{tabular}}
\end{table}

% Finally, we use $\alpha$ for a consistency check. The product of $\alpha$ and $d$ is a parameter derived from the PL fitting, and its value is $ \alpha d \approx 6$. We can calculate the penetration depth, $d$, of the $780\ \mathrm{nm}$ source used for the photoconductive current measurement using $d =\sqrt{\mu\ \mathrm{k_B T}\ (300\ \mathrm{ns}) / q}$ and the corresponding $\tau$. We find that $d \approx 2\ \mathrm{\mu m}$. This would then lead to $\alpha_\mathrm{780\ nm} \approx 30,000\ \mathrm{cm^{-1}}$ which is quite close to the first-principles computed value of $10,000\ \mathrm{cm^{-1}}$ \cite{Yuan2024DiscoveryAbsorber}. 

{\color{black} Finally, we can use the absolute PL intensity to obtain an order-of-magnitude estimate of the photoluminescence quantum yield (PLQY). Although we did not use an integrating sphere, we can assume that the PL emission radiates from the samples in a hemispherical manner, and, using the objective lens's collection solid angle, we can estimate the PLQY. Detailed calculations are presented in \hyperref[sec:s3]{SI-Section 3.5}. We find that for GaAs-Wafer, $\mathrm{PLQY}_\mathrm{{GaAs}} \approx 0.76\ (\pm\ 0.05)\%$ for a power density of $185\ \mathrm{W/cm^2}$. For BCP-Crystal $\mathrm{PLQY_{BCP}} \approx  0.20\ (\pm\ 0.01)\%$ at a power density of $410\ \mathrm{W/cm^2}$. BCP-Crystal’s PLQY estimate is notable because it implies that, even without surface passivation, BCP-Crystal has a PLQY in the same order of magnitude as that of the best CdTe devices with surface passivation, which have a PLQY of $10^{-3}$, i.e., $\mathrm{PLQY_{CdTe}} \approx  0.1\ \%$ \cite{Bowman2024SpatiallyFilms, Scarpulla2023CdTe-basedProspects, Mallick2023Arsenic-DopedEfficiency}.}

\textbf{Deep Defect Recombination Rate}. The impressive band-to-band PL intensity, implied $V_\mathrm{OC}$, and carrier lifetime of BCP compared to GaAs could originate from BCP’s favorable intrinsic defect properties, tolerance to extrinsic impurities, lower rate of surface recombination, or a combination of all these. To better probe and compare the bulk properties of our BCP and GaAs samples, we study their point defects and nonradiative recombination rates using first-principles calculations. 

Point defects, whether they have shallow or deep levels, can reduce optical emission and carrier transport \cite{Hammer2025BridgingPhotovoltaics,McCluskey2012DopantsSemiconductors}. Shallow defects have charge-state transition levels typically within a few $k_\mathrm{B}T$ of the band edges, and can induce scattering events by trapping and de-trapping charge carriers, thereby lowering carrier mobility \cite{V.N.1991NonradiativeSemiconductors, Mishra2002AlGaN/GaNApplications, Freysoldt2009FullyCalculations, Freysoldt2014First-principlesSolids, Kuciauskas2023WhyCells, Scajev2025AsDopedCells}. On the other hand, deep defects create levels near the middle of the band gap, and are particularly detrimental to the carrier lifetime because they can act as SRH nonradiative recombination centers, causing photocurrent and $V_\mathrm{OC}$ loss \cite{Simoen2015AnalyticalDetection, Liu2020EmergingStrategies, Shukla2021CarrierCuInGaS2}. The lower the SRH recombination rate, the less detrimental the deep defect would be \cite{Wurfel2005PhysicsCells}. The SRH rate for the defect-assisted recombination is given by \cite{Shockley1952StatisticsElectrons, Das2020WhatTheory}:
\begin{equation}
R_{SRH} = \frac{np-n_i^2}{\frac{1}{NC_p}(n+n_1)+\frac{1}{NC_n}(p+p_1)},
\label{eq:8}
\end{equation}
where $N$ is the defect concentration. $C_n$ and $C_p$ are electron and hole capture coefficients, $n$ and $p$ are electron and hole concentrations, and $n_1$ and $p_1$ are electron and hole concentrations when the Fermi level is at the defect level.

\begin{figure}[!t]
    \centering
    \includegraphics[width=1.0\linewidth]{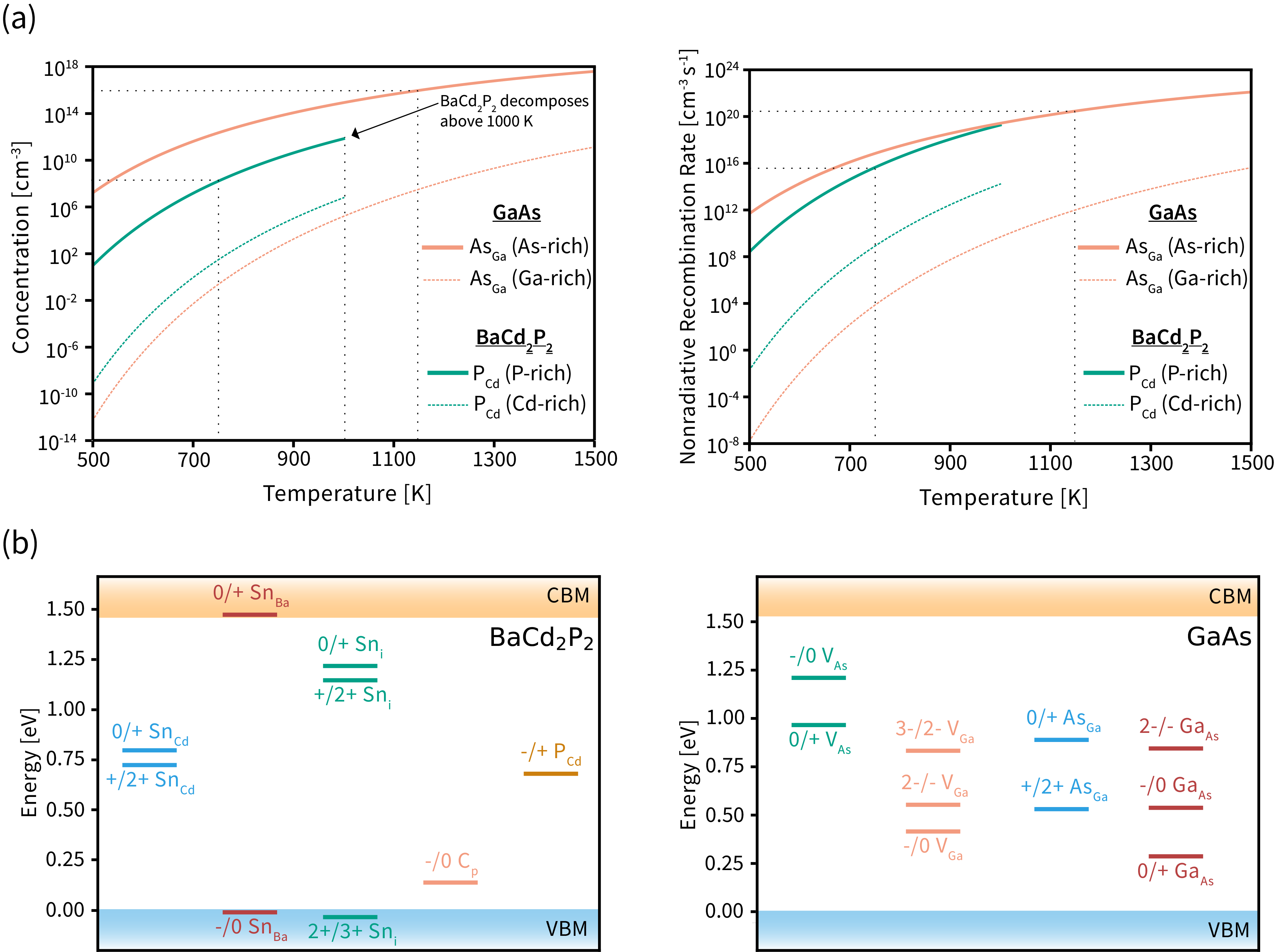}
    \caption{(a) The evolution of $\mathrm{P_{Cd}}$ and $\mathrm{As_{Ga}}$ deep defect concentrations assuming they were frozen at temperatures ranging from 500 K to the respective synthesis temperatures of BCP and GaAs (left), and the resulting SRH recombination rate (right). We note that GaAs is prepared under As-rich conditions, and as a result, the Ga-rich results are not typically realized. (b) The charge transition levels of point defects in BCP (left) and GaAs (right). Sn and C impurity defects are explored in BCP in addition to the intrinsic P on Cd antisite. Aside from As on Ga antisite, As vacancy, Ga vacancy, and Ga on As antisite defects are presented for GaAs.}
    \label{fig:5}
\end{figure}

{\color{black} Combinations of first-principles defect calculations, and experiment have been used to identify defect-related recombination and performance-limiting mechanisms in photovoltaic absorbers, including CdTe \cite{Ma2013DependenceCalculations} and, more recently, trigonal selenium \cite{Kavanagh2025IntrinsicPhotovoltaics}. Here, w}e use density functional theory (DFT) with the hybrid functional of Heyd-Scuseria-Ernzerhof (HSE) \cite{Kohn1965Self-ConsistentEffects, Heyd2003HybridPotential} to calculate the SRH rate of our BCP-Crystal and GaAs-Wafer samples. Details of the first-principles calculations are provided in \hyperref[sec:s5]{SI-Section 5}. We first examine intrinsic point defects in both materials, followed by a discussion of impurities. In our previous work, we showed that the $\mathrm{P_{Cd}}$ antisite is the dominant nonradiative recombination center in BCP, with $(-/0)$ and $(0/+)$ transitions deep in the band gap \cite{Yuan2024DiscoveryAbsorber}. 

\hyperref[fig:5]{Figure 5b} shows the charge transition levels of intrinsic defects in GaAs. The dominant deep defect in GaAs, often referred to as EL2, has been widely attributed to the $\mathrm{As_{Ga}}$ antisite \cite{Kaminska1993ChapterGaAs, SammyKayali1997GaAsApplications}. EL2 has been proposed both as an isolated defect \cite{Skowronski1985MetastabilityDefect, Weber1982IdentificationGaAs, Kaminska1985IdentificationDefect} and as part of a defect complex \cite{vonBardeleben1986IdentificationGaAs, Wager1987AtomicGaAs, Baraff1988ModellingEL2}. Some have attributed it to the $(+/2+)\ \mathrm{As_{Ga}}$ transition and others to the $(0/+)\ \mathrm{As_{Ga}}$ transition \cite{Chadi1988MetastabilityGaAs, Dabrowski1988TheoreticalEL2, Wampler2015TemperatureArsenide, Kaminska1987EL2GaAs}. The ionization energy of EL2 has been experimentally reported to be $0.825\ \mathrm{eV}$ away from the conduction band minimum (CBM) using deep-level transient spectroscopy (DLTS) \cite{Kaminska1987EL2GaAs, Martin1980KeyGaAs}. This is equivalent to $0.595\ \mathrm{eV}$ above the valance band minimum (VBM) when considering $E_\mathrm{g}\ (300\ \mathrm{K})=1.42\ \mathrm{eV}$. From the HSE configuration coordinate and formation energy diagrams of $\mathrm{As_{Ga}}$ (\hyperref[fig:s7]{Figure S7}), we see that the ionization energy of $(+/2+)\ \mathrm{As_{Ga}}$ is $0.53\ \mathrm{eV}$ above the VBM, and is a better match to the DLTS reported value of EL2 than $(0/+)\ \mathrm{As_{Ga}}$, whose calculated ionization energy is $0.89\ \mathrm{eV}$. 
%This result agrees with other recent HSE calculations \cite{KhangHoang2025First-principlesGaAs}. 
Furthermore, this HSE calculated ionization energy of $(+/2+)\ \mathrm{As_{Ga}}$ agrees well with photo-electron paramagnetic resonance studies which have reported a value of $0.52\ \mathrm{eV}$ above the VBM for the capture of a hole by $\mathrm{As_{Ga}^{+1}}$ \cite{Weber1982IdentificationGaAs, Kaminska1987EL2GaAs, Elliott1984IdentificationGaAs}. Additionally, we calculated the classical energy barrier for nonradiative recombination, $E_\mathrm{B}$, and found that $E_\mathrm{B} = 0.24\ \mathrm{eV}$ for the capture of hole by $\mathrm{As_{Ga}^{+1}}$ in the $(+/2+)$ transition (\hyperref[fig:s7]{Figure S7}, \hyperref[table:s4]{Table S4}). This value agrees more with the experimentally determined $E_\mathrm{B}=0.3\ \mathrm{eV}$ for EL2 \cite{Kaminska1993ChapterGaAs, Kaminska1987EL2GaAs} than the calculated $E_\mathrm{B}=1.4\ \mathrm{eV}$ for electron and hole capture during the $(0/+)\ \mathrm{As_{Ga}}$ transition. These results suggest that $(+/2+)\ \mathrm{As_{Ga}}$ is a better match to the experimentally observed properties of EL2 than $(0/+)\ \mathrm{As_{Ga}}$.

\begin{table}[t]
    \centering

    {\color{black} \small
    \setlength{\tabcolsep}{0.7pt}
    \caption{Charge transition levels (eV) for native point defects in GaAs from selected literature. Transition energies are referenced to the VBM. Each column header lists the reference, defect supercell size, and the DFT method (for HSE, the mixing parameter $\alpha$ is shown). The Supercell size of 64/216 denotes the case in which defect geometries were relaxed in a 64-atom cell and then embedded in a 216-atom cell. $G_0W_0$@HSE denotes one-shot GW calculations performed on an HSE starting point.}
    
    \begin{tabular}{cc|c|c|c|c|c|c}
    \\
    \hline
    \hline
    && This Work & \cite{Fluckey2026PointStudy} & \cite{Komsa2012IntrinsicCalculations, Komsa2012ComparisonFunctionals} & \cite{Chen2017AccuracySolids} & \cite{Komsa2011IdentificationFunctionals}  & \cite{Komsa2011AssessingGaAs} \\
    &                               & HSE             & HSE               & HSE             & $G_0W_0$@HSE  & HSE               & HSE                  \\
    &                               & ($\alpha=0.28$) & ($\alpha=0.26$)   & ($\alpha=0.35$) &               & ($\alpha=0.35$)   & ($\alpha=0.395$)     \\
\textbf{Defect} && 512 atoms       & 216 atoms         & 64 atoms        & 64 atoms      & 64/216 atoms      & 64/216 atoms         \\
\hline
    $\mathrm{As_{Ga}}$     &(+2/+)  & 0.53            & 0.48              & 0.63            & 0.5           & 0.56              & 0.57                 \\   
                           &(+/0)   & 0.89            & 0.67              & 0.98            & 0.8           & 0.90              & 0.97                 \\
    \hline
    $\mathrm{Ga_{As}}$     &(+2/+)  & -               & 0.06              & 0.03            & -             & -                 & -                    \\   
                           &(+/0)   & 0.29            & 0.19              & 0.3             & -             & -                 & -                    \\
                           &(0/-)   & 0.53            & 0.64              & 0.4             & -             & -                 & -                    \\
                           &(-/-2)  & 0.84            & 0.69              & 0.8             & -             & -                 & -                    \\
    
    \hline
    \end{tabular}}
\end{table}

As detailed in \hyperref[sec:s5]{SI-Section 5}, we determine the concentration of the deep defects, then quantum-mechanically compute the carrier capture coefficients using Fermi’s golden rule within the static coupling formalism and a 1D configuration coordinate model \cite{Turiansky2021Nonrad:Principles,Alkauskas2014First-principlesEmission, Stoneham2001TheorySolids,Li2019EffectiveSections}. The first step in calculating the SRH recombination rate is determining the defect concentration. Defect concentrations depend on {\color{black}formation energies, which in turn depend on the elemental chemical potentials, i.e., on the growth conditions. Accordingly, we calculate the defect formation energies for the chemical potential ranges in which BCP and GaAs are stable (\hyperref[fig:s9]{Figure S9}). We use the formation energies to determine the defect concentrations and SRH recombination rates for both materials.} In addition, defect concentrations depend on temperature. It is often assumed that the defects formed at synthesis temperatures are kinetically locked in during cooling \cite{Buckeridge2019EquilibriumEnergy, Squires2025GuidelinesCrystals, Sasaki1999LowOxides}. However, atoms are usually mobile even until $\sim500\ \mathrm{K}$ when samples are not quenched \cite{Abe2011IntrinsicResults}. In \hyperref[fig:5]{Figure 5a} we present how the defect concentrations would evolve when considering the deep defects in BCP and GaAs are frozen at different temperatures ranging from $500\ \mathrm{K}$ to the synthesis temperatures (1000 K for BCP and 1510 K for GaAs). 

The EL2 defect concentration has been reported to be $\approx$$10^{16}\ \mathrm{cm}^{-3}$ for GaAs grown using Liquid Encapsulated Czochralski (LEC), such as our GaAs-Wafer \cite{Rudolph1999BulkOverview}. From the dotted lines in \hyperref[fig:5]{Figure 5a}, we see that this concentration is achieved at $1160\ \mathrm{K}$, i.e., when assuming the defects are frozen at $3/4\ \mathrm{th}$ of the synthesis temperature of GaAs under As-rich conditions. This finding is consistent with the fact that GaAs is grown in As-rich conditions, maintained through a high As flux for LEC, and most other techniques of GaAs growth \cite{Rudolph1999BulkOverview, Liu1994MechanismGaAs, Zhang1991ChemicalSelf-diffusion}. Applying the same concept to BCP, we consider the defect concentrations at 750 K, $3/4\ \mathrm{th}$, the synthesis temperature of BCP (BCP dissociates above 1000 K). Using these temperatures, we find that $\mathrm{P_{Cd}}$ in BCP results in a SRH rate ranging from $8.05 \times 10^{8}$ to $4.36 \times 10^{15}\ \mathrm{{cm^{-3}\ s^{-1}}}$ when considering a typical photo-injection level of $\Delta n = 10^{14}\ \mathrm{cm^{-3}}$ (this $\Delta n$ is between the $\Delta n \approx10^{12}\ \mathrm{cm}^{-3}$ seen for BCP under photoconductive current measurements and $\Delta n \approx 10^{17}\ \mathrm{cm}^{-3}$ seen for PL measurements). On the other hand, the SRH rate resulting from $\mathrm{As_{Ga}}$ in GaAs is $3.45 \times 10^{20}\ \mathrm{{cm^{-3}\ s^{-1}}}$ for the As-rich conditions used in our GaAs-Wafer sample growth when considering $\Delta n =10^{14}\ \mathrm{cm^{-3}}$ and using $1160\ \mathrm{K}$. In fact, \hyperref[fig:5]{Figure 5a} shows that under the typical As-rich growth, $\mathrm{As_{Ga}}$ in GaAs has a higher SRH rate than $\mathrm{P_{Cd}}$ in BCP across temperatures ranging from $500\ \mathrm{K}$ to the respective synthesis temperatures of BCP and GaAs. This lower SRH recombination rate illustrates the high PV potential of BCP, considering that GaAs solar cells have demonstrated high power conversion efficiencies \cite{Hautzinger2025SynthesisApplications, Cotal2009IIIVPhotovoltaics, Yoon2010GaAsAssemblies}. {\color{black} The defect concentrations under varying chemical potentials and the capture coefficients used to calculate the SRH rates, and the $E_\mathrm{B}$ and $E_\mathrm{0}$ values are presented in Tables \hyperref[table:s4]{S4}, \hyperref[table:s5]{S5}, and \hyperref[table:s6]{S6}. The raw DFT relaxation files are available in our repository.}

The carrier lifetimes of BCP and GaAs that would result from their intrinsic deep defects are longer than what we observe in the experimental carrier lifetime measurements; in BCP, $\mathrm{P_{Cd}}$ would result in $2.3\times10^{-2}\ \mathrm{s}$ and in GaAs, $\mathrm{As_{Ga}}$ would result in $2.9\times10^{-6}\ \mathrm{s}$. There could be several reasons for the difference between the measured and computed ($\mathrm{P_{Cd}}$ and $\mathrm{As_{Ga}}$ limited) lifetimes, such as assumptions made in the temperature at which the defects get frozen, fast surface recombination channels, impurity related deep defect centers with high SRH nonradiative recombination rates, and more. Here, we explore the possibility that impurity-related deep traps are the cause of the lower measured lifetime. Accordingly, we examine potential deep defects arising from impurities introduced during synthesis and in the raw materials used for BCP. First, we explore impurities from the synthesis of BCP. Because BCP-Crystal was grown in Sn flux, we explore the possibility of having Sn impurity-related defects (\hyperref[fig:5]{Figure 5b}). We see that the $\mathrm{Sn_{Cd}}$ substitutional defect introduces deep levels. However, BCP-Powder samples, which were synthesized without Sn flux, had PL emissions in the same order as BCP-Crystal, suggesting that $\mathrm{Sn_{Cd}}$ is likely not a prominent nonradiative recombination pathway. Moreover, the carbonized silica ampoule possibly introduces C and Si. We tested the possibility of C substituting P in BCP and found that this defect would only form a shallow level; other C related defects have yet to be explored.

Next, considering the raw materials, we observe that the top impurities in the Ba source used to prepare BCP are Ca, Sr, and Cu \cite{ThermoFisherScientificInc.BariumAnalysis}. Cu stands out among these, given that transition metal impurities typically form deep-level defects in semiconductors, and Cu is a well-known deep-level defect center in GaAs \cite{Sze2006PhysicsDevices}. The Cu concentration in the Ba source is 32 ppm. However, recent works show that the $\mathrm{Cu_{Cd}}$ substitutional and Cu interstitial defects do not form deep levels \cite{Coban2025ExploringAbsorbers, Coban2024ExploringAbsorbers}. Furthermore, the P raw material has 157 ppm Fe, which is another known impurity that can result in deep defects in GaAs and other semiconductors \cite{Sze2006PhysicsDevices}; the high optoelectronic performance of BCP despite the elevated levels of Fe impurity in the precursors is impressive. Similarly, for GaAs, there are several deep defect candidates that could limit its lifetime, including oxygen impurities and defect complexes \cite{Hoang2025First-principlesGaAs}.
% We performed HSE defect calculations to see if Cu would introduce any deep-level defects in BCP, specifically the $\mathrm{Cu_{Cd}}$ substitutional defect, as shown in \hyperref[fig:5]{Figure 5b}, we found that the $\mathrm{Cu_{Cd}}$ substitutional defect only has shallow transitions.

In summary, we demonstrated that BCP is a defect-tolerant analogue of GaAs. Both are direct-gap materials with a ${0\ \mathrm{K}}$ band gap of $1.51\ \mathrm{eV}$. Room-temperature PL showed that BCP powder exhibits bright band-to-band emissions, whereas GaAs powder, synthesized using similar solid-state reactions, is primarily dominated by defect emissions. Even compared to a ground, prime-grade GaAs wafer, BCP powder exhibits a higher band-to-band PL emission peak intensity, despite having much lower precursor purity. These findings suggest that BCP has a higher tolerance to impurities and a lower surface recombination rate than GaAs. We prepared a crystal of BCP using Sn flux and found that it has a higher implied $V_\mathrm{OC}$ than a pristine wafer of GaAs for incident optical power densities near{\color{black} $100 \ \mathrm{\mathrm{W/cm^2}}$. The smaller temperature dependence of the BCP band gap compared to that of GaAs partly contributes to the higher $V_\mathrm{OC}$ of BCP. We were able to estimate the PLQY of the BCP crystal using our PL photon flux calibration, and found it to be $\sim0.2\%$ at a power density of $410\ \mathrm{W/cm^2}$; this value is in the same order as that of the GaAs wafer, for which the PLQY is $\sim 0.76\%$ at a power density of $185\ \mathrm{W/cm^2}$.} Moreover, photoconductive current measurements showed $\tau \geq 300 \ \mathrm{ns}$ for the BCP crystal while $\tau \approx 5 \ \mathrm{ns}$ for the GaAs wafer at the power density of $0.164\ \mathrm{W/cm^2}$ of a 780 nm excitation source (3.2 Suns equivalent). Finally, we used first-principles calculations to show that the dominant intrinsic deep defect in BCP has a lower SRH recombination rate than that of GaAs under typical growth conditions. BCP is also tolerant to some extrinsic impurities, such as C and Sn, introduced during its synthesis, and transition metal impurities coming from its raw materials. The Ba source used for BCP contains Cu, which only forms a shallow level in BCP. The P source contains high levels of Fe, which is typically detrimental in GaAs, but BCP exhibits impressive optoelectronic performance despite the high Fe content. Such impurity tolerance makes BCP a promising counterpart of GaAs for further improving PV performance at lower cost.

\section*{Data availability}
The supporting information text is available in the appendices, and {\color{black} the DFT data generated in this study, including relaxed defect structures, and the scripts for the post-processing the calculations, the }spontaneous emission PL fitting model, and the SRH rate calculations are all available on GitHub at \href{https://github.com/gideon116/BaCd2P2-aPIT-counterpart-of-GaAs-for-PVs}{github.com/gideon116/BaCd2P2-aPIT-counterpart-of-GaAs-for-PVs}.

\section*{Contributions}
G.K. and Z.Y. worked on the first-principles computations; M.R.H. and K.K. worked on solid-state synthesis and crystal growth; G.K. worked on the PL and electrical characterizations with guidance on data analyses from J.L.; G.K., Z.Y., and J.L. prepared the manuscript, with inputs from G.H., K.K., G.L.E., and D.P.F.

\section*{Acknowledgments}
This work was supported by the U.S. Department of Energy, Office of Science, Basic Energy Sciences, Division of Materials Science and Engineering, Physical Behavior of Materials program under award number DE-SC0023509. This research used resources of the National Energy Research Scientific Computing Center (NERSC), a DOE Office of Science User Facility supported by the Office of Science of the U.S. Department of Energy under contract no. DE-AC02-05CH11231 using NERSC award BES-ERCAP0020966.

% \begingroup
%   \sloppy
%   \printbibliography[
%     title={References}]
% \endgroup

% \end{refsection}

% \begin{refsection}

\clearpage
\section*{Supporting Information}
\setcounter{figure}{0}
\renewcommand{\thefigure}{S\arabic{figure}}
\setcounter{equation}{0}
\renewcommand{\theequation}{S\arabic{equation}}
\setcounter{table}{0}
\renewcommand{\thetable}{S\arabic{table}}

\tableofcontents
\bigskip
\hrule
\bigskip

\section{Solid State Synthesis and Sn Flux Growth}\label{sec:s1}

The synthesis of the BCP powders was carried out using elemental reactants in a stoichiometric 1:2:2 ratio (Ba:Cd:P). Dendritic Ba (99.9\%, Alfa Aesar), Cd powder (99.5\%, Sigma Aldrich) or cadmium shots (99.95\%, Alfa Aesar), and red phosphorus powder (98.90\%, Alfa Aesar) were used for the reactions. Crystals were synthesized in different batches using elements from different vendors over the course of this study. The mixtures were placed in carbonized $9/11\ \mathrm{mm}$ ID/OD silica ampoules inside an argon-filled glovebox. The ampoules were evacuated and sealed with a hydrogen-oxygen torch, then heated in a muffle furnace to over $800\mathrm{^{\circ}C}$ for $8-10\ \mathrm{hours}$. After annealing at this temperature for $48\ \mathrm{hours}$, the furnace was turned off to allow the samples to cool. The ampoules were then opened under ambient conditions, and the products were ground using agate mortars. To improve phase purity, the ground powders were reannealed at a similar temperature for another $48\ \mathrm{hours}$ in evacuated carbonized silica ampoules. The final products were used for further analysis.

\begin{figure}[!t]
    \centering
    \includegraphics[width=0.6\linewidth]{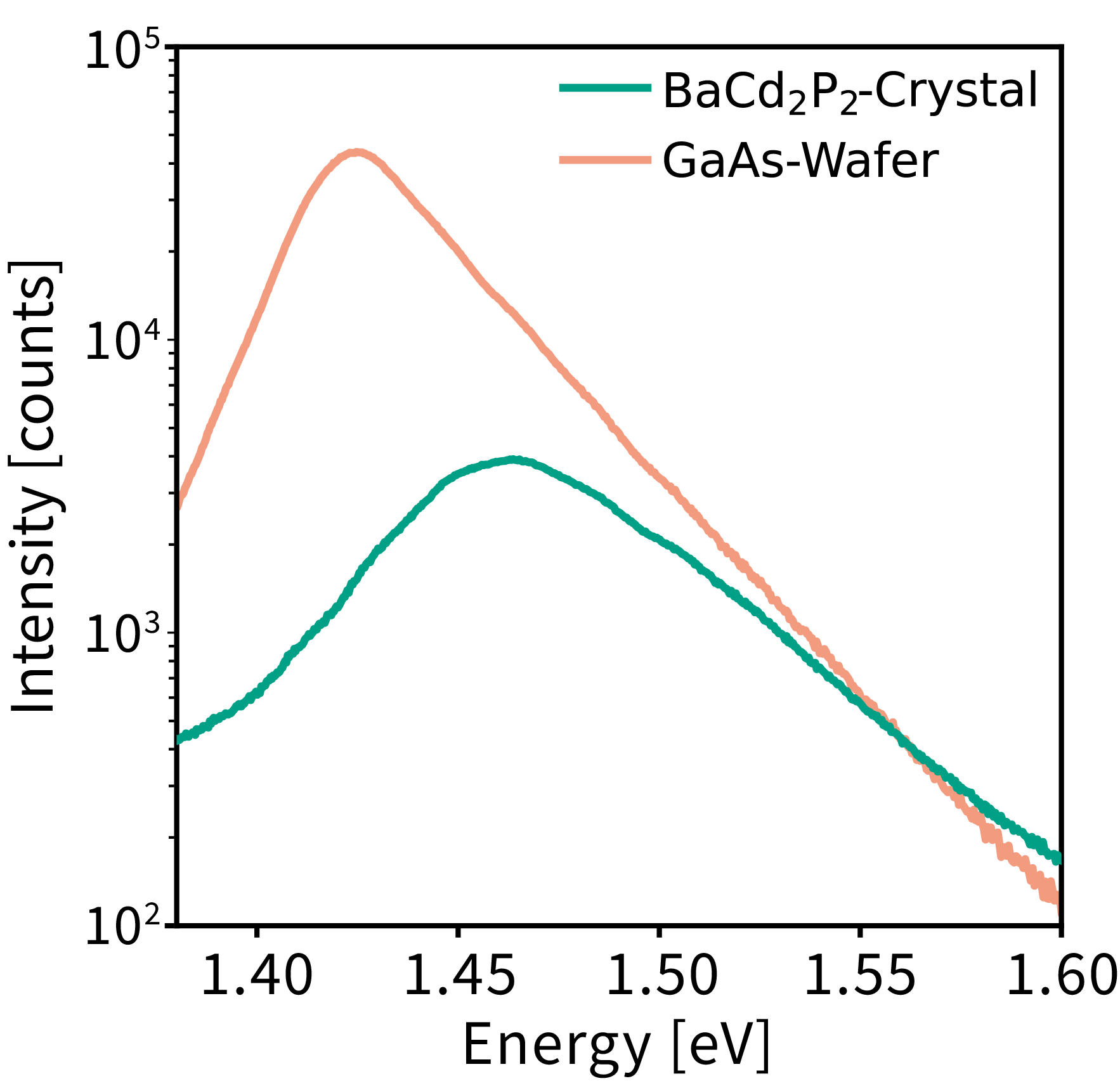}
    \caption{PL spectra of BCP-Crystal and single crystalline GaAs-Wafer at $298\ \mathrm{K}$. A $532\ \mathrm{nm}$ laser was used with an incident excitation power density of $141.5\ \mathrm{W/cm^2}$.}
    \label{fig:s1}
\end{figure}

For Sn flux crystal growth, the respective elements having the same sources and purities as mentioned were loaded into the carbonized $9/11\ \mathrm{mm}$ ID/OD silica ampoules along with metallic Sn (99.5\%, Alfa Aesar) in a 1:2:2:20 Ba:Cd:P:Sn molar ratio in an argon-filled glovebox. A layer of silica wool was then placed on top of the reaction mixture such that it does not come into contact with the elements. Chips of silica were then placed on top of the silica wool, and the ampoules were then evacuated and sealed using a hydrogen-oxygen torch. After being placed in a muffle furnace in an upright position (silica chips at the top, elements at the bottom), the ampoules were heated to $900\mathrm{^{\circ}C}$ for $9\ \mathrm{hours}$ and annealed at that temperature for $6\ \mathrm{hours}$ followed by cooling to $600 \mathrm{^{\circ}C}$ over $75\ \mathrm{hours}$, after which were quickly taken out, inverted and rapidly centrifuged while hot. Doing so allowed the molten Sn flux to be collected at the bottom of the ampoule along with the silica chips, while the crystals either remained stuck to the ampoule wall or caught in the silica wool. The ampoules were then cooled and opened in ambient conditions, and the crystals were picked out. In some cases, depending on how much Sn was stuck to the crystals, centrifugation was carried out a second time in a different ampoule in a similar fashion to yield cleaner crystals, which were further etched to remove residual Sn from the surface. 

\begin{figure}[!t]
    \centering
    \includegraphics[width=0.6\linewidth]{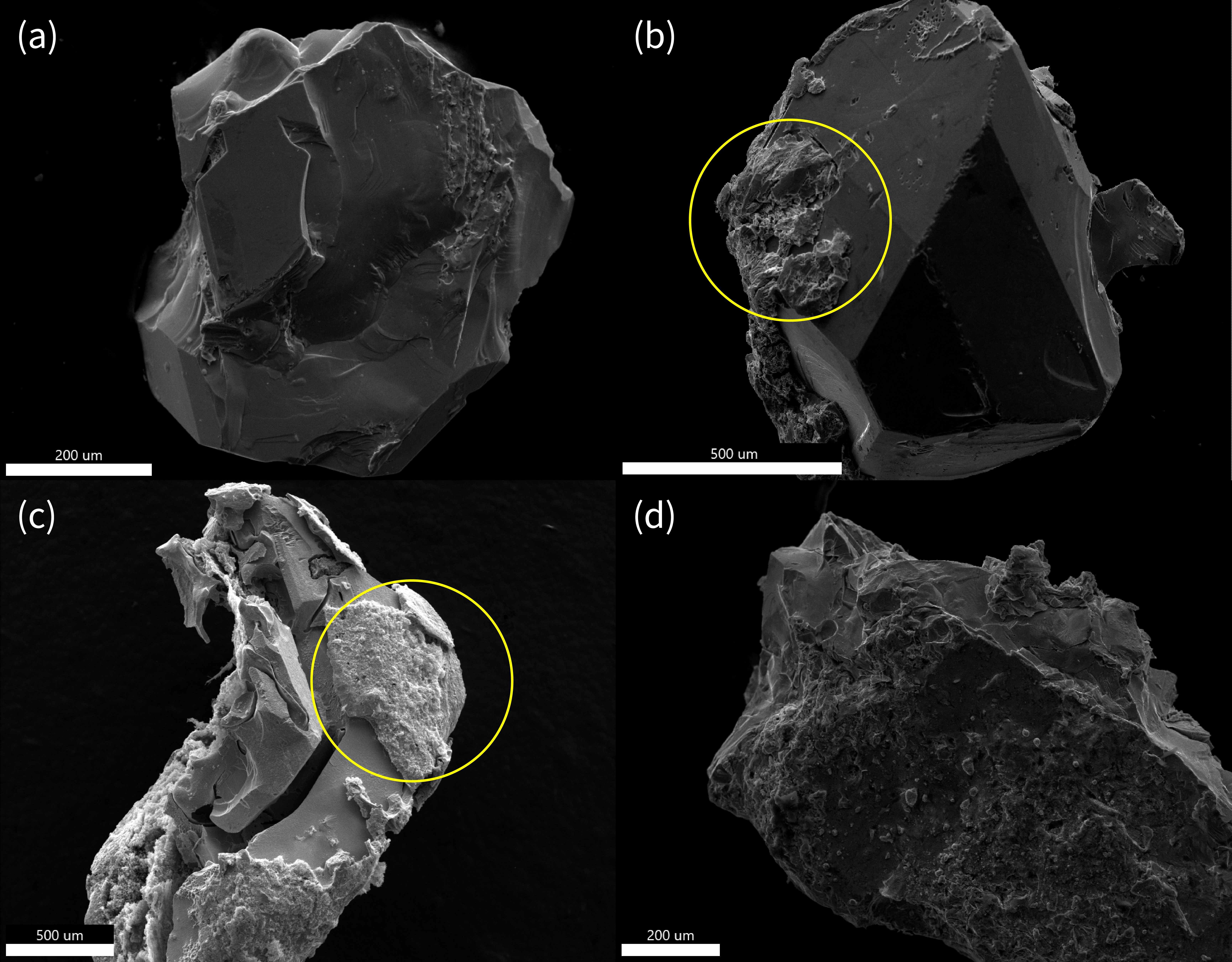}
    \caption{SEM images of BCP crystals synthesized using Sn flux. (a) and (b) were etched with weak HCl to remove residual Sn, while (c) and (d) were treated using strong HCL and NaOH solutions, respectively. The yellow circles point to regions on the crystals' surface that have residual Sn.}
    \label{fig:s2}
\end{figure}

Energy dispersive X-ray spectroscopy (EDS) and single crystal X-ray diffraction (SCXRD) experiments verified the crystals to be BCP. The amount of residual Sn on the surface varies from one crystal to another. For instance, crystals (a) and (b) in \hyperref[fig:s2]{Figure S2} were etched using dilute HCl to remove Sn. EDS showed that the rough/amorphous regions, circled in yellow, are Sn-rich. We observe that crystal (a) contains little to no Sn, whereas (b) contains some residual Sn. In addition to dilute HCl, (c) and (d) were treated with strong HCl and NaOH, respectively. From (c), we see that strong HCl etches the crystal faster than it etches Sn, so using strong HCl to remove residual Sn only works if the amount of Sn is significantly smaller than the crystal size. EDS of (d) shows little to no residual Sn remaining. For our optoelectronic measurements, we selected clean crystals, like (a).

\section{Varshni Fitting}\label{sec:s2}

Fitting the blueshift in the PL peak of BCP using the Varshni equation we find that \(E_{g} = \ 1.5116\ \mathrm{eV} - \frac{{5.3150\times {10}^{-4}\ \mathrm{\frac{eV}{K}}\times T}^{2}}{T\  + \ 806.75\ \mathrm{K}}\ \mathrm{eV}\), where $T$ is the temperature (in units of $\mathrm{K}$) and $1.512\ \mathrm{eV}$ is the band gap at $0\ \mathrm{K}$ (\hyperref[fig:1]{Figure 1b}) \cite{Varshni1967TemperatureSemiconductors}. This value is in good agreement with the HSE06-calculated $0\ \mathrm{K}$ band gap of  $1.45\ \mathrm{eV}$ \cite{Yuan2024DiscoveryAbsorber}. Similarly, fitting the blueshift of GaAs using the Varshni equation we find that \(E_{g} = 1.5133\ \mathrm{eV} - \frac{{5.3642\times10^{- 4}\ \mathrm{\frac{eV}{K}}\times T}^{2}}{T\  + \ 249.72\ \mathrm{K}}\ \mathrm{eV}\) (\hyperref[fig:1]{Figure 1b}). The derived band gap of $1.513\ \mathrm{eV}$ at $0\ \mathrm{K}$ for GaAs is almost identical to that of BCP and agrees well with the accepted GaAs band gap of $1.512\ \mathrm{eV}$ \cite{Levinshtein1996HandbookParameters, Blakemore1982SemiconductingArsenide}. 

\hyperref[table:s1]{Table S1} shows the evolution of the FWHM of the PL peaks with temperature. As the temperature decreases below $\mathrm{173\ K}$, emissions from radiative defects start to emerge, making it difficult to isolate and measure the FWHM of the band-to-band peaks.

\begin{table}[!t]
\centering
\caption{The ratio of the full width half maximum of the band-to-band PL peaks of BCP and GaAs to $\mathrm{1.8\ k_B T}$. A value of $1$ is the ideal case for a single emission, considering a parabolic band model. The ratio artificially increases with decreasing temperature as radiative defect emissions become prevalent.\\}
\label{table:s1}
        \begin{tabular}{ccc}
        \toprule
        Temperature ($\mathrm{K}$)&  BCP $\mathrm{\left( FWHM / 1.8\ k_B T \right)}$& GaAs $\mathrm{\left( FWHM / 1.8\ k_B T \right)}$\\
        \midrule
        296&  1.53& 0.93\\
        263&  -& 1.03\\
        253&  1.53& -\\
        233&  -& 1.08\\
        213&  1.56& -\\
        203& -&1.17\\
        173&  1.75& 1.28\\
        % 143&  -& 1.46\\
        % 133& 1.98&-\\
        % 113& -&1.78\\
        % 78&  2.91& 2.35\\
        \bottomrule
    \end{tabular}
\end{table}

\section{PL Fitting Derivation}\label{sec:s3}

Here, we provide the full derivation for the spontaneous emission model used to fit the PL. We provide the code for performing this fitting on \href{https://github.com/gideon116/BaCd2P2-aPIT-counterpart-of-GaAs-for-PVs}{GitHub} repository associated with this paper. We also provide approximations to this model that are less computationally demanding. Finally, we provide an overview of how we calibrate our micro-PL setup to determine the absolute PL intensity.

\subsection{Deriving the Spontaneous Emission Model}

\begin{figure}[!t]
    \centering
    \includegraphics[width=0.7\linewidth]{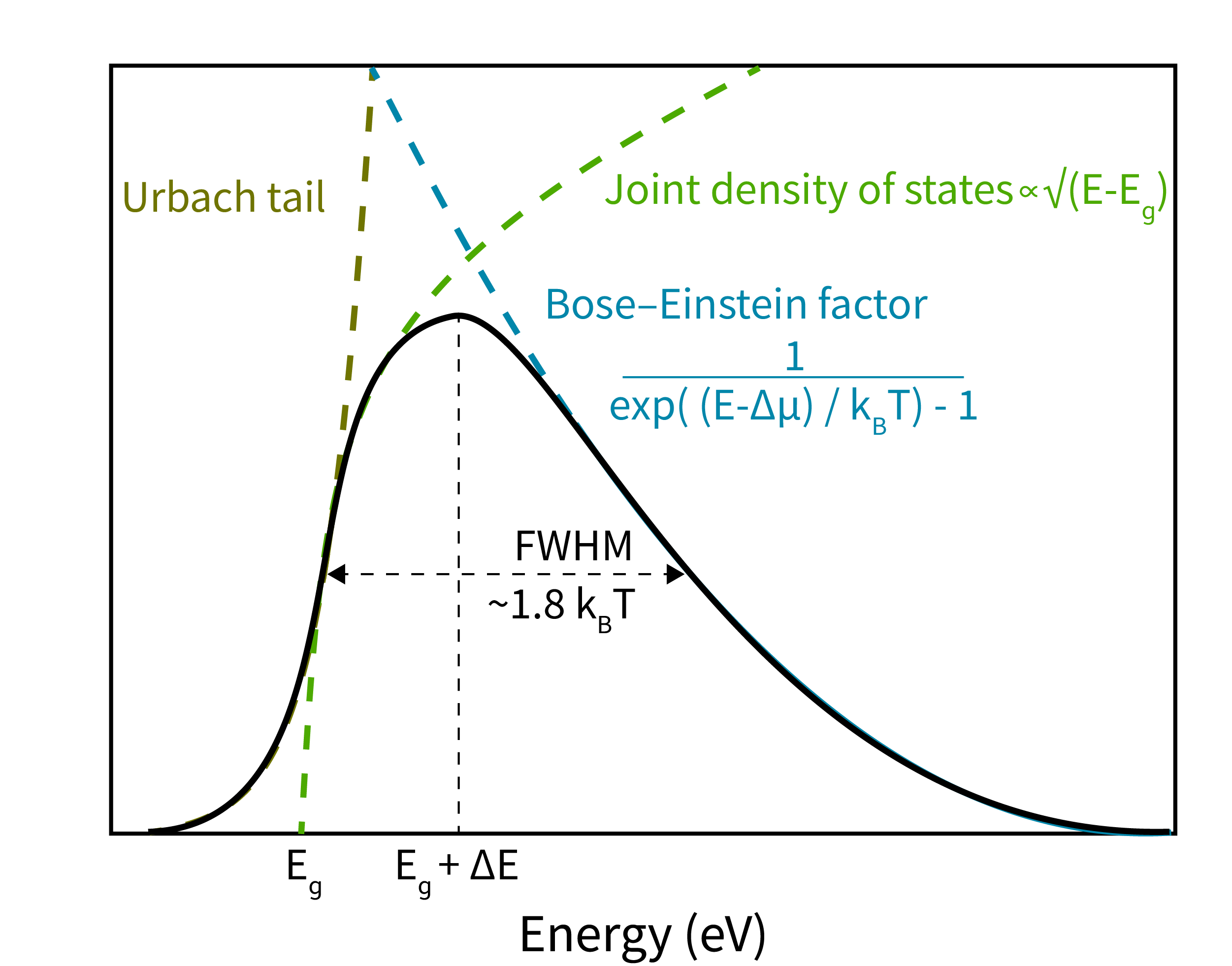}
    \caption{Schematic of the band-to-band PL lineshape near the absorption edge. The spectrum follows \hyperref[eq:s1]{Equation S1}. Above the edge, it follows $a(E)\propto\sqrt{E-E_g}$ (green), below the edge, it is shaped by the Urbach tail (olive). The blue dashed curve shows the Bose–Einstein factor with $\Delta \mu$. The PL maximum occurs at $E_g+\Delta E$, where $\Delta E\sim\mathcal{O}(k_{B}T)$. For a single direct transition under low injection, the FWHM is $\sim$$1.8\ \mathrm{k_B T}$; very strong injection or with multiple transitions can lead to a deviation from this principle.}
    \label{fig:s3}
\end{figure}

A schematic of the components of a PL spectrum is presented in \hyperref[fig:s3]{Figure S3}. To understand the different sections that make up the spectrum, we begin with the van Roosbroeck and Shockley \cite{vanRoosbroeck1954Photon-RadiativeGermanium} equation:
\begin{equation}
I_{PL} = \frac{2 \pi}{c^{2}h^{3}}\frac{E^{2}}{\exp{\left( \frac{E - \mathrm{\Delta}\mu}{k_BT} \right) - 1}}a(E)
\label{eq:s1}
\end{equation}

where \(I_{PL}\) is the absolute intensity in \(\mathrm{\frac{Photons}{m^{2} \ s \ eV}}\) and \(a(E)\) is the absorptivity given by
\begin{equation}
a(E) = 1 - \exp\left( - \alpha(E) \cdot d \right)
\label{eq:s2}
\end{equation}
here, $d$ is the characteristic absorption length and $\alpha(E)$ is the absorption coefficient which is proportional to the joint density of states (DOS) assuming ideal parabolic bands, i.e., \(\frac{1}{2 \pi^{2}} \left( \frac{2 m_{e}^{*}}{\hbar^{2}} \right)^{\frac{3}{2}} \sqrt{E - E_{g}}\), where $m_{e}^{*}$ is the electron effective mass. Thus:
\begin{equation}
{\alpha(E) = \alpha}_{0} \ \rho(E) = \alpha_{0}  \sqrt{E - E_{g}}
\label{eq:s3}
\end{equation}
To account for low energy band tails, we convolve \(\alpha(E)\) with \(T(E)\) where \(T(E)\) is \cite{Katahara2014Quasi-FermiPhotoluminescence}:
\begin{equation}
T(E) = N \ \exp\left( - \left| \frac{E\  - E_{g}}{E_{U}} \right|^{\theta} \right)
\label{eq:s4}
\end{equation}
where \(E_{U}\) is the Urbach energy and \(N\) is the value that ensures the integral of the convolution from \(- \infty\) to \(\infty\) is 1. To find \(N\):
\[1 = \int_{- \infty}^{\infty}{N \ \exp\left( - \left| \frac{E\  - E_{g}}{E_{U}} \right|^{\theta} \right)\ dE}\]
\begin{equation}
N = \frac{1}{{2 E}_{U} \Gamma\left( 1 + \frac{1}{\theta} \right)}
\label{eq:s5}
\end{equation}
So now we have:
\[{\alpha(E) = \alpha}_{0} \sqrt{E - E_{g}} \times T(E)\]
\begin{equation}
{\alpha(E) = \alpha}_{0} \frac{1}{{2 E}_{U} \Gamma\left( 1 + \frac{1}{\theta} \right)} \int_{- \infty}^{E}{\exp{\left( - \left| \frac{x\ - E_{g}}{E_{U}} \right|^{\theta} \right)  \sqrt{E - x - E_{g}}}\ dx}
\label{eq:s6}
\end{equation}
To account for the fact that already excited carriers cannot take part in absorption, we scale $\alpha(E)$ by $f_{h}\left( E, E_{F}^{h} \right) - f_{e}\left( E, E_{F}^{e} \right)$, which is difference in the Fermi-Dirac distributions of  the quasi-Fermi levels of electrons ($E_{F}^e$) and holes $(E_{F}^h$) :
\begin{equation}
f_{e}\left(E, E_F^{e} \right) - \ f_h\left(E, E_F^h \right) = \left(1 - \frac{1}{\exp{\left( \frac{E - E_F^h}{k_BT} \right) + 1}} \right) - \frac{1}{\exp{\left( \frac{E - E_F^e}{k_BT} \right) + 1}}
\label{eq:s7}
\end{equation}
Finally, we have:
\begin{equation}
a(E) = 1 - exp( - \alpha(E) \cdot d \cdot (f_{h}\left( E,E_{F}^{h} \right) - f_{e}\left( E,E_{F}^{e} \right)))
\label{eq:s8}
\end{equation}
{\color{black} \begin{equation}
I_{PL} = \frac{2 \pi}{c^{2}h^{3}}\frac{E^{2}}{\exp{\left( \frac{E - \mathrm{\Delta}\mu}{k_BT} \right) - 1}} a(E)
\label{eq:s9}
\end{equation}}

\subsection{Fitting PL Spectra with Spontaneous Emission Model}

{\color{black} When fitting the PL data, we do not fit individual PL spectra independently. Instead, for each material we perform a global fit to multiple PL spectra ($N\ge 8$, where $N$ is the number of spectra) collected over a range of excitation power densities, while constraining the material specific quantities (e.g., $E_g$, $T$, $E_U$, and $\alpha(E)\cdot d$) to be shared across all PL spectra. Only the power-dependent $\Delta \mu$ is allowed to vary from spectrum to spectrum. This reduces the risk of overfitting due to over-parameterization compared with fitting individual spectra independently. Now, while one could in principle fit $\Delta \mu$ (by fitting $E_f^e$ and $E_f^h$, and using $\Delta \mu = E_f^e -E_f^h$) independently for each spectrum (for each power level), we instead fit the photogenerated carrier density, $\Delta n$, to reduce the number of fitting parameters. We know that $E_{F}^{e}$ and $E_{F}^{h}$ are given by}
\begin{equation}
E_{F}^{e}\ = E_C - k_B T \ \ln\left(\frac{N_C}{n} \right), \ 
E_{F}^{h}\ = E_V + k_B T\  \ln\left(\frac{N_V}{p} \right)
\label{eq:s10}
\end{equation}
$N_C$ and  $N_V$ are the effective density of states in the conduction and valence bands, respectively, and are given by:
\begin{equation}
N_{C} = 2\left( \frac{2 \pi  m_{e}^{*}  \  k_B T}{h^{2}} \right)^{\frac{3}{2}}, \ N_{V} = 2\left( \frac{2 \pi m_{h}^{*} \ k_B T}{h^{2}} \right)^{\frac{3}{2}}
\label{eq:s11}
\end{equation}
where $m_e^*$ and $m_h^*$ are the electron effective masses. {\color{black} $n$ and $p$ are the electron and hole concentrations, and given by $n = n_0 + \mathrm{\Delta}n$ and $p = p_0 + \mathrm{\Delta}n$. For an undoped material such as our BCP samples, $n_0$ and $p_0$ are the intrinsic electron and hole concentrations, respectively (i.e., assuming $n_0 = p_0 = n_i$), and can be obtained using}
\begin{equation}
n_0 = N_C \  \exp\left( - \frac{E_{g} - E_{F}}{k_B \cdot T} \right), \ 
p_0 = N_V \ \exp\left( - \frac{E_{F}}{k_B \cdot T} \right)
\label{eq:s12}
\end{equation}
where, $E_{F}$ is the equilibrium Fermi level. {\color{black} We can then use these $n_0$ and $p_0$ values in $n = n_0 + \mathrm{\Delta}n$ and $p = p_0 + \mathrm{\Delta}n$ then calculate $E_F^e$ and $E_F^h$.

On the other hand, an n-type doped material, such as our GaAs wafer, with a doping concentration $N_D$, is treated differently. Because $n_0 \approx N_D$, we can use the relation}
\begin{equation}
n_{i}^{2} = n_{0}p_{0} = {N_{C} \cdot N}_{V} \cdot \exp\left( - \frac{E_{g}}{k_B \cdot T} \right)
\label{eq:s13}
\end{equation}
{\color{black} to find $p_0 \approx \frac{n_i^2}{N_D}$. Upon carrier injection of $\Delta n$, we get $n = N_D + \mathrm{\Delta}n$ and $p = \frac{n_i^2}{N_D} + \mathrm{\Delta}n$. We can then use these values to calculate $E_F^e$ and $E_F^h$.}

Thus, the power dependent PL spectra can be fit using $\mathrm{\Delta}n$ as {\color{black} the power dependent fitting parameter, and $E_g$, $T$, $E_U$, $\alpha(E) \cdot d$, $m_{e}^{*}$, and $m_{h}^{*}$ (with $\theta$ fixed at 1) as global material parameters.

For BCP, the PL is fitted to two band-to-band emissions and therefore includes two sets of $E_g$ and $E_U$.} Even when there are multiple emissions arising from closely located band minima, as is the case with BCP, there will be a single $\mathrm{\Delta}\mu$. This is because carrier quasi-equilibrium must be ensured, and carrier thermalization between the multiple conduction band minima is faster than radiative recombination. As a result, we iteratively solve the charged balance equation until charge neutrality, $n(\Delta \mu)-p(\Delta \mu) = n_0-p_0$, is satisfied to within $10^{-9}\ \mathrm{cm^{-3}}$ (see \href{https://github.com/gideon116/BaCd2P2-aPIT-counterpart-of-GaAs-for-PVs}{github.com/gideon116/BaCd2P2-aPIT-counterpart-of-GaAs-for-PVs} for fitting details). In addition, we use the DFT derived values $m_e^* \approx 0.15\ m_0$ and $m_h^* \approx 0.5\ m_0$ from \cite{Yuan2024DiscoveryAbsorber}.

{\color{black} The parameters extracted from the spontaneous emission model fit provide information about the material, including carrier lifetime, $n_i$, $\Delta \mu$, $N_C$, $N_V$, and more. For instance, the fitted temperature of BCP is $315\  \mathrm{K}$, for which solving for the charge-neutrality condition when considering the dominant point defect in BCP, as described in \hyperref[eq:s5]{SI Section 5}, yields $n_{i} \approx 2 \times 10^7\ \mathrm{cm^{- 3}}$.

\subsection{Simplifying the Model for PL Fitting}

While it is best to use the rigorous absorptivity equation \hyperref[eq:s8]{Equation S8}, one can also use a simplified model that makes certain approximations to improve computational efficiency. In our calculations, we adopted the full form of the absorptivity model in \hyperref[eq:s8]{Equation S8} to calculate the PL emission using \hyperref[eq:s9]{Equation S9} (referred to as ``Full Model"). Below, we discuss three simplifications (Cases I-III) that give a good approximation of the PL (referred to as ``Simple Model"):}

\textbf{Case I.} If we assume injected carriers are much more than material doping and the electron and hole effective masses are nearly equal, then \(E_{F}^{e}\) = \(E_{F}^{h}\) and:
\[f_{h}\left( E,E_{F}^{h} \right) - f_{e}\left( E,E_{F}^{e} \right) = \left( 1 - \frac{1}{\exp{\left( \frac{E - \frac{\mathrm{\Delta}\mu}{2}}{k_BT} \right) + 1}} \right) - \frac{1}{\exp{\left( \frac{E - \frac{\mathrm{\Delta}\mu}{2}}{k_BT} \right) + 1}}\]
\begin{equation}
f_{h}\left( E,E_{F}^{h} \right) - f_{e}\left( E,E_{F}^{e} \right) = 1 - \frac{2}{\exp{\left( \frac{E - \mathrm{\Delta}\mu/2}{k_BT} \right) + 1}}
\label{eq:s14}
\end{equation}

\textbf{Case II.} If we assume the material is moderately to heavily n-doped, $\Delta\mu$ $\sim$ \(E_{F}^{h}\):
\[f_h\left(E, E_F^h \right) - f_e\left(E,E_F^e \right) = \left(1 - \frac{1}{\exp{\left(\frac{E-\Delta\mu}{k_BT} \right) + 1}} \right) - \frac{1}{\exp{\left(\frac{E}{k_BT} \right) + 1}}\]
\begin{equation}
f_{h}\left( E,E_{F}^{h} \right) - f_{e}\left( E,E_{F}^{e} \right) = 1 - \frac{\exp\left( \frac{E - \mathrm{\Delta}\mu}{k_{B}T} \right) + \exp{\left( \frac{E}{k_BT} \right) + 2}}{\exp{\left( \frac{2E - \mathrm{\Delta}\mu}{k_BT} \right) + \exp\left( \frac{E}{k_{B}T} \right)} + \exp\left( \frac{E - \mathrm{\Delta}\mu}{k_BT} \right) + 1}
\label{eq:s15}
\end{equation}

\textbf{Case III.} When only interested in modeling the PL line shape and not the absolute intensity (and hence \(\mathrm{\Delta}\mu\)), we can use the relation \cite{Merrick2007PhotoluminescenceNonparabolicity, Latkowska2013TemperatureGap, Ullrich1991OpticalFilms, Bebb1972ChapterTheory}:
\begin{equation}
I_{PL}(E)
= \left\{
  \begin{array}{ll}
    C\,D\,\exp\ \Bigl(\frac{\sigma}{k_{B}T}(E - E_{T})\Bigr)\,f(E),
      & E < E_{T}, \\[6pt]
    C\,\sqrt{E - E_{g}}\,f(E),
      & E \ge E_{T}.
  \end{array}
\right.
\label{eq:s16}
\end{equation}
where \(I_{PL}\) corresponds to the normalized PL intensity,  \(\sigma\) describes the slope of the Urbach tail, \(C\) is the joint DOS prefactor (left as a fit parameter), and \(f(E)\) is the Fermi-Dirac occupation factor. When \(E < E_{T}\), the function considers sub-band gap states. \(E_{T}\) is found by equating the two equations at \(E = E_{T}\), giving \(E_{T} = E_{g} + D^{2}\). A smooth transition between the two equations at \(E_{T}\) is ensued using \(D\), which is determined by equating the derivatives of the two equations at \(E = E_{T}\), giving \(D = \frac{\sqrt{k_{B}T}}{2 \cdot \sigma}\). This equation holds under low injection levels where the quasi-Fermi levels are more than \(k_{B}T\) away from their respective band edges \cite{Yu2010FundamentalsSemiconductors, Kohn1957ShallowGermanium}. In the low-injection regime, the shape of the normalized PL spectrum remains unchanged until Auger recombination becomes noticeable; this means that while $\mathrm{\Delta}n$ is increasing, the shape of the normalized PL remains the same up to a certain point. Thus, we can neglect the absolute PL intensity and use \hyperref[eq:s16]{Equation S16} for a much faster calculation, which accurately predicts the PL shape.

{\color{black} We performed PL fitting with both the Simple Model and the Full Model. We used the simplification in Case I for BCP and the simplification in Case II for GaAs. Similar to the Full Model fitting, the Simple Models are global fits of multiple power-dependent PL spectra, with material constants such as $E_g$ and $E_U$ shared across all spectra. We compare the extracted parameters from the Simple Models and Full models in Tables \hyperref[table:s2]{S2} and \hyperref[table:s3]{S3} for BCP and GaAs, respectively. The extracted $\Delta\mu$ values (and therefore implied $V_\mathrm{OC}$) and the key material parameters $E_g$ and $E_U$ are nearly unchanged between the two models; this indicates that our conclusions are insensitive to the generalization of the Full Model.}

\begin{table}[!ht]
    \centering
    {\color{black}
    \caption{Comparison of fitted parameters for BCP using the full PL model and the simplified model. BCP is fit as a convolution of two band-to-band emission components (1 and 2), with $(E_{g1},E_{U1})$ and $(E_{g2},E_{U2})$ describing the corresponding band-gap and Urbach-energy parameters. Reported values are the median with subscripts/superscripts giving the empirical 95\% interval (2.5th-97.5th percentiles) obtained from an ensemble of randomized multi-start fits. In each trial, we also perturb the photon-to-count calibration by $\pm 60$ photons/count to account for calibration uncertainty; only fits with $R^2>0.998$ are retained.}
    
    \label{table:s2}
    \renewcommand{\arraystretch}{1.75}
    \begin{tabular}{l l l}
    \\
        \hline
        \hline
        Parameter & Full Model & Simplified Model   \\
        \hline
        $E_{U1}$ (eV) & $0.010_{-0.001}^{+0.001}$ & $0.011_{-0.001}^{+0.001}$ \\
        \hline
        $E_{U2}$ (eV) & $0.019_{-0.001}^{+0.001}$ & $0.013_{-0.001}^{+0.002}$ \\
        \hline
        $E_{g1}$ (eV) & $1.440_{-0.003}^{+0.001}$  & $1.441_{-0.001}^{+0.002}$  \\
        \hline
        $E_{g2}$ (eV) & $1.459_{-0.002}^{+0.001}$  & $1.456_{-0.001}^{+0.010}$  \\
        \hline
        
        $\Delta \mu_1$  (eV) & $1.189_{-0.006}^{+0.009}$ & $1.200_{-0.010}^{+0.019}$ \\
        \hline
        $\Delta \mu_2$  (eV) & $1.215_{-0.004}^{+0.010}$ & $1.227_{-0.010}^{+0.019}$ \\
        \hline
        $\Delta \mu_3$  (eV) & $1.230_{-0.003}^{+0.011}$ & $1.243_{-0.010}^{+0.019}$ \\
        \hline
        $\Delta \mu_4$  (eV) & $1.240_{-0.002}^{+0.012}$ & $1.254_{-0.010}^{+0.019}$ \\
        \hline
        $\Delta \mu_5$  (eV) & $1.248_{-0.002}^{+0.012}$ & $1.262_{-0.010}^{+0.019}$ \\
        \hline
        $\Delta \mu_6$  (eV) & $1.254_{-0.002}^{+0.013}$ & $1.268_{-0.010}^{+0.019}$ \\
        \hline
        $\Delta \mu_7$  (eV) & $1.260_{-0.001}^{+0.013}$ & $1.274_{-0.010}^{+0.019}$ \\
        \hline
        $\Delta \mu_8$  (eV) & $1.265_{-0.001}^{+0.013}$ & $1.280_{-0.010}^{+0.019}$ \\
        \hline
        $\Delta \mu_9$  (eV) & $1.270_{-0.001}^{+0.014}$ & $1.285_{-0.009}^{+0.019}$ \\
        \hline
        $\Delta \mu_10$ (eV) & $1.275_{-0.001}^{+0.014}$ & $1.289_{-0.009}^{+0.018}$ \\
    \hline
    \end{tabular}}
\end{table}

\begin{table}[!ht]
    \centering
    {\color{black}
    \caption{Comparison of fitted parameters for GaAs using the full PL model and the simplified model. Reported values are the median with subscripts/superscripts giving the empirical 95\% interval (2.5th-97.5th percentiles) obtained from an ensemble of randomized multi-start fits. In each trial, we also perturb the photon-to-count calibration by $\pm 60$ photons/count to account for calibration uncertainty; only fits with $R^2>0.998$ are retained.}
    
    \label{table:s3}
    \renewcommand{\arraystretch}{1.75}
    \begin{tabular}{l l l}
    \\
        \hline
        \hline
        Parameter & Full Model & Simplified Model \\
        \hline
        $E_U$ (eV) & $0.010_{-0.001}^{+0.002}$ & $0.009_{-0.001}^{+0.001}$   \\
        \hline
        $E_g$ (eV) & $1.429_{-0.001}^{+0.013}$ & $1.441_{-0.009}^{+0.036}$   \\
        \hline

        $\Delta \mu_1$ (eV) & $1.132_{-0.001}^{+0.001}$ & $1.133_{-0.001}^{+0.001}$ \\
        \hline
        $\Delta \mu_2$ (eV) & $1.153_{-0.001}^{+0.001}$ & $1.153_{-0.001}^{+0.001}$ \\
        \hline
        $\Delta \mu_3$ (eV) & $1.164_{-0.001}^{+0.001}$ & $1.164_{-0.001}^{+0.001}$ \\
        \hline
        $\Delta \mu_4$ (eV) & $1.171_{-0.001}^{+0.001}$ & $1.172_{-0.001}^{+0.001}$ \\
        \hline
        $\Delta \mu_5$ (eV) & $1.177_{-0.001}^{+0.001}$ & $1.178_{-0.001}^{+0.001}$ \\
        \hline
        $\Delta \mu_6$ (eV) & $1.183_{-0.001}^{+0.001}$ & $1.183_{-0.001}^{+0.001}$ \\
        \hline
        $\Delta \mu_7$ (eV) & $1.188_{-0.001}^{+0.001}$ & $1.187_{-0.001}^{+0.001}$ \\
        \hline
        $\Delta \mu_8$ (eV) & $1.191_{-0.001}^{+0.001}$ & $1.191_{-0.001}^{+0.001}$ \\
        \hline
        $\Delta \mu_9$ (eV) & $1.194_{-0.001}^{+0.001}$ & $1.194_{-0.001}^{+0.001}$ \\
    \hline
    \end{tabular}}
\end{table}

\subsection{PL Photon to Count Calibration}

{\color{black} To obtain the absolute PL intensity, we calibrated the PL system to correlate the detector reading to the number of photons. We did so through the following approach:

\begin{enumerate}
\def\labelenumi{\arabic{enumi}.}
\item Shine a laser of a given wavelength on a wafer of Si.
\item  Calculate the expected intensity of the Raman peak (photon flux) for the supplied laser power, while considering the optical efficiency of the system, and detector quantum efficiency.
\item  Correlate the expected photon flux to the detector reading (in counts/s) to obtain a photon-to-count correspondence.
\item  Check values by repeating steps 1-3 using excitation lasers with varying wavelengths.
\end{enumerate}

To begin, we calculate the Raman intensity ($I$), treating the sample as a semi-infinite, strongly absorbing, backscattering sample using Beer–Lambert depth integration.
\begin{equation}
I = \frac{P_\mathrm{laser}} {E_\mathrm{laser}} \cdot S\cdot (\alpha_\mathrm{laser} + \alpha_\mathrm{raman})^{-1} \cdot \Omega_\mathrm{eff}
\label{eq:s17}
\end{equation}
\begin{equation}
I = \frac{P_\mathrm{laser}} {E_\mathrm{laser}} \cdot \frac{d\sigma}{d\Omega}\cdot N \cdot (\alpha_\mathrm{laser} + \alpha_\mathrm{raman})^{-1} \cdot \Omega_\mathrm{eff}
\label{eq:s18}
\end{equation}
where $P_\mathrm{laser}$ is the fraction of the source laser power ($P_\mathrm{source}$) on the sample surface after accounting for the optical efficiency ($\eta_\mathrm{opt}$) of the system and the ND filter ($\mathrm{ND_{filter}}$). $P_\mathrm{laser}$ is given by
\begin{equation}
P_\mathrm{laser} = P_\mathrm{source} \cdot \eta_\mathrm{opt} \cdot \mathrm{ND_{filter}}
\label{eq:s19}
\end{equation}
\begin{equation}
P_\mathrm{laser} = 100\ \mathrm{mW} \cdot 29\% \cdot 100\% = 0.029\ \mathrm{J/s}
\label{eq:s20}
\end{equation}
$E_\mathrm{laser}$ is incident laser energy, and here is $E_\mathrm{laser}=E_\mathrm{532nm} = 3.73\times10^{-19}\ \mathrm{J/photon}$. $N$ is number density for Si atoms where $N_\mathrm{Si} = 5.0\times10^{22}\ \mathrm{cm^{-3}}$. $\alpha_\mathrm{laser}$ and $\alpha_\mathrm{raman}$ are the absorption coefficients of Si at the laser (532 nm) and Raman (547 nm) wavelengths where $\alpha_\mathrm{532nm} = 1.0 \times 10^4\  \mathrm{cm^{-1}}$ and $\alpha_\mathrm{547nm} = 9.3 \times 10^3\  \mathrm{cm^{-1}}$. The collection solid angle for our 0.55 NA objective is $\Omega_\mathrm{eff} = 2 \pi\ (1-\mathrm{cos}(\mathrm{arcsin}(0.55))) = 6.28\times (1-0.835) = 1.04\ \mathrm{steradians}$.

$\frac{d\sigma}{d\Omega} = \frac{\sigma_\mathrm{RS}}{4 \pi}$, where $\sigma_\mathrm{RS}$ is the total Raman cross section of Si, which is laser intensity dependent. Table 1 in \cite{Aggarwal2011MeasurementSilicon} shows that $\sigma_\mathrm{RS}^{1.92\ \mathrm{eV}} = 7.9\pm5.5\times10^{-27}\ \mathrm{cm^2}$ and $\sigma_\mathrm{RS}^{2.41\ \mathrm{eV}} = 5.2\pm2.3\times10^{-26}\ \mathrm{cm^2}$. Because $7.9\pm5.5\times10^{-27}\ \mathrm{cm^2} < \sigma_\mathrm{RS} < 5.2\pm2.3\times10^{-26}\ \mathrm{cm^2}$, we use the theoretical polarizability and scattering values in \cite{Grimsditch1980AbsoluteSilicon} and \cite{Renucci1975ResonantSilicon} to make the approximation: %1.92<2.41 eV.
\begin{equation}
\frac{d\sigma}{d\Omega} = \frac{\sigma_\mathrm{RS}}{4 \pi} = 1.0\times 10^{-26}\ / 4 \pi = 7.96\times 10^{-28}\ \mathrm{cm^2/steradians}
\label{eq:s21}
\end{equation}
It then follows that:
\begin{equation}
I = 1.65 \times 10^8\ \mathrm{photons/s}
\label{eq:s22}
\end{equation}

Finally, to perform the calibration, we use the Raman intensity ($I_\mathrm{meas}$), i.e., the baseline-subtracted line integral after applying wavelength-dependent detector quantum-efficiency and objective-lens transmission corrections.
\begin{equation}
    I/I_\mathrm{meas} = \frac{1.65 \times 10^8\ \mathrm{photons/s}}{1.63\pm0.09\times 10^5\ \mathrm{counts/s}} = 1.01\ \pm 0.06\times 10^3\ \mathrm{photons/count}
\label{eq:s23}
\end{equation}
%The propagated error for the function  f(x) = 1/x is given by the formula  e_f = e_x/x^2.

To test our approach, we use Si Raman spectra collected with 532 nm and 633 nm lasers, with the remaining settings kept the same. We calculate the expected ratio of Raman intensities, $I_\mathrm{532nm}/I_\mathrm{633nm}$, and compare it to the experimentally measured ratio.
\begin{equation}
I_\mathrm{532nm}/I_\mathrm{633nm} = \frac{(d\sigma/d\Omega)_\mathrm{532nm}}{(d\sigma/d\Omega)_\mathrm{633nm}}\cdot \frac{P_\mathrm{532nm}}{P_\mathrm{633nm}}\cdot \frac{E_\mathrm{633nm}}{E_\mathrm{532nm}} \cdot \frac{(\alpha_\mathrm{633nm} + \alpha_\mathrm{655nm})}{(\alpha_\mathrm{532nm} + \alpha_\mathrm{547nm})}
\label{eq:s24}
\end{equation}
\begin{equation}
\begin{split}
\frac{I_{\mathrm{532nm}}}{I_{\mathrm{633nm}}} = \,\, & \frac{1.0 \times 10^{-26}\ \mathrm{cm^2}}{7.9 \pm 5.5 \times 10^{-27}\ \mathrm{cm^2}} \cdot \frac{29\ \mathrm{mW}}{10\ \mathrm{mW}} \cdot \frac{3.14\times10^{-19}\ \mathrm{\frac{J}{photon}}}{3.73\times10^{-19}\ \mathrm{\frac{J}{photon}}} \\
& \cdot \frac{(4.6 \times 10^3\  \mathrm{cm^{-1}} + 4.0 \times 10^3\  \mathrm{cm^{-1}})}{(1.0 \times 10^4\  \mathrm{cm^{-1}} + 9.3 \times 10^3\  \mathrm{cm^{-1}})}
\end{split}
\label{eq:s25}
\end{equation}
\begin{equation}
I_\mathrm{Si,532nm}/I_\mathrm{Si,633nm} = 1.4\pm0.9
\label{eq:s26}
\end{equation}
The detector quantum efficiency and objective lens transmission adjusted integrated Raman intensities are $1.63\pm0.09\times 10^5$ counts/s and $1.51\pm0.05\times 10^5$ counts/s for the 532 nm and 633 nm lasers, respectively; thus $I_\mathrm{meas,532nm}/I_\mathrm{meas,633nm} = 1.08\pm 0.07$. This value is within the error of the expected ratio of Raman intensities for 532 nm and 633 nm excitation lasers.

\subsection{PLQY OoM Estimate}

The absolute PL calibration enables us to estimate the PLQY as follows:
\begin{equation}
P_\mathrm{laser}^\mathrm{{GaAs}} = P_\mathrm{source} \cdot \eta_\mathrm{opt} \cdot \mathrm{ND_{filter}^\mathrm{{GaAs}}}
\label{eq:s27}
\end{equation}

The nominal laser power $P_\mathrm{source}$ is 45 mW, thus
\begin{equation}
P_\mathrm{laser}^\mathrm{{GaAs}} = 45\ \mathrm{mW} \cdot 0.29 \cdot 0.1\% = 1.305 \times 10^{-5}\ \mathrm{W}
\label{eq:s28}
\end{equation}
When considering that the laser spot diameter is $3\ \mathrm{\mu m}$, as determined in \hyperref[sec:s3]{SI-Section 3.6}, we see that the corresponding power density is $185\ \mathrm{W/cm^2}$.
\begin{equation}
\Phi_\mathrm{laser}^\mathrm{{GaAs}} = P_\mathrm{{laser}}^\mathrm{{GaAs}} \cdot \frac{\lambda} {hc}\cdot (1-R) %abs_coe * d
\label{eq:s29}
\end{equation}
\begin{equation}
\Phi_\mathrm{laser}^\mathrm{{GaAs}} = 1.305 \times 10^{-5} \cdot \frac{532 \times 10^{-9}} {6.626 \times 10^{-34} \cdot 3 \times 10^{8}}\cdot (1-0.324) %abs_coe * d
\label{eq:s30}
\end{equation}
\begin{equation}
\Phi_\mathrm{laser}^\mathrm{{GaAs}} = 2.36 \times 10^{13}\ \mathrm{photons/s}
\label{eq:s31}
\end{equation}

The detector quantum efficiency adjusted integrated PL intensity of GaAs (after multiplying with $2 \pi/\Omega_\mathrm{eff}$ to estimate hemispherical external emission and accounting for wavelength-dependent transmission of the objective lens) is:
\begin{equation}
\Phi_\mathrm{PL}^\mathrm{{GaAs}} = 1.80\pm 0.11 \times 10^{11}\ \mathrm{photons/s}
\label{eq:s32}
\end{equation}

Therefore, PLQY can be estimated using:
\begin{equation}
\mathrm{PLQY}_\mathrm{{GaAs}} = \Phi_\mathrm{PL}^\mathrm{{GaAs}}/\Phi_\mathrm{laser}^\mathrm{{GaAs}} = 0.0076\pm0.0005 = 0.76\pm0.05\%
\label{eq:s33}
\end{equation}

For BCP, the detector quantum efficiency adjusted integrated PL intensity is:
\begin{equation}
\Phi_\mathrm{PL}^\mathrm{{BCP}} = 1.05\pm0.06 \times 10^{11}\ \mathrm{photons/s}
\label{eq:s34}
\end{equation}
The nominal laser power used was 100 mW and considering $\eta_\mathrm{opt} = 0.29$ and $\mathrm{ND_{filter}^\mathrm{{BCP}}=0.1\%}$:
\begin{equation}
\Phi_\mathrm{laser}^\mathrm{BCP} = 5.25 \times 10^{13}\ \mathrm{photons/s}
\label{eq:s35}
\end{equation}
This corresponds to a power density of $410\ \mathrm{W/cm^2}$.
\begin{equation}
\mathrm{PLQY_{BCP}} = \frac{1.05\pm0.06 \times 10^{11}\ \mathrm{photons/s}}{5.25 \times 10^{13}\ \mathrm{photons/s}} = 0.0020\pm0.0001 = 0.20\pm0.01\%
\label{eq:s36}
\end{equation}}

\subsection{PL Power Density Estimation}

\begin{figure}
    \centering
    \includegraphics[width=0.8\linewidth]{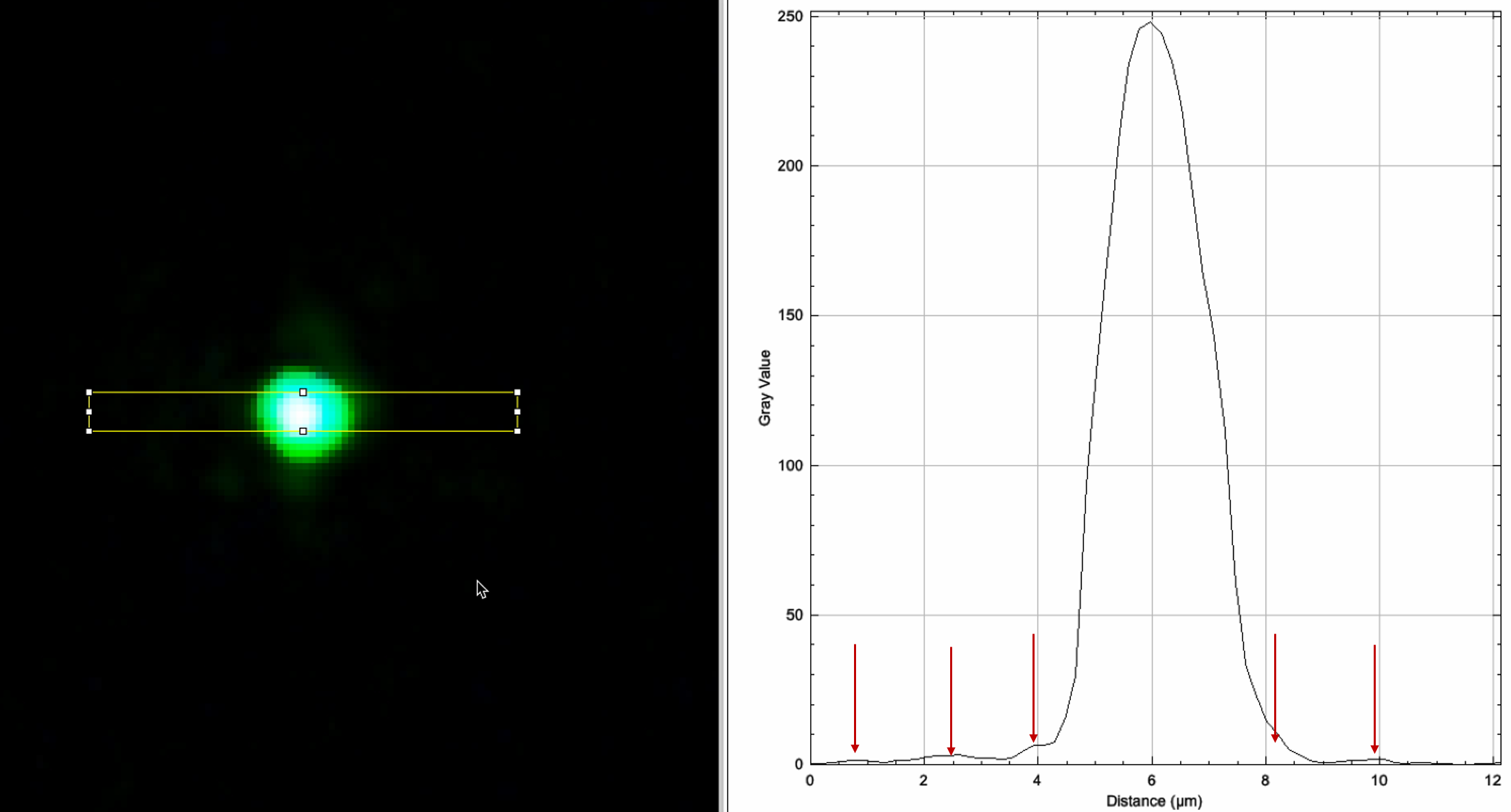}
    \caption{Laser spot characterization used to determine excitation power density. The image shows the central Airy disk with $\mathrm{FWHM} \approx 3\ \mathrm{\mu m}$. In the corresponding line profile, the primary peak is centered near $x \approx 6\ \mathrm{\mu m}$, with first order diffraction shoulders at $x \approx 4\ \mathrm{\mu m}$ and $x \approx 8\ \mathrm{\mu m}$, and second order diffraction shoulders at $x \approx 2\ \mathrm{\mu m}$ and $x \approx 10\ \mathrm{\mu m}$. The width of the central Airy disk provides a lower limit estimate of the optical excitation area, assuming negligible scattered light.}
    \label{fig:s4}
\end{figure}

\begin{figure}
    \centering
    \includegraphics[width=0.5\linewidth]{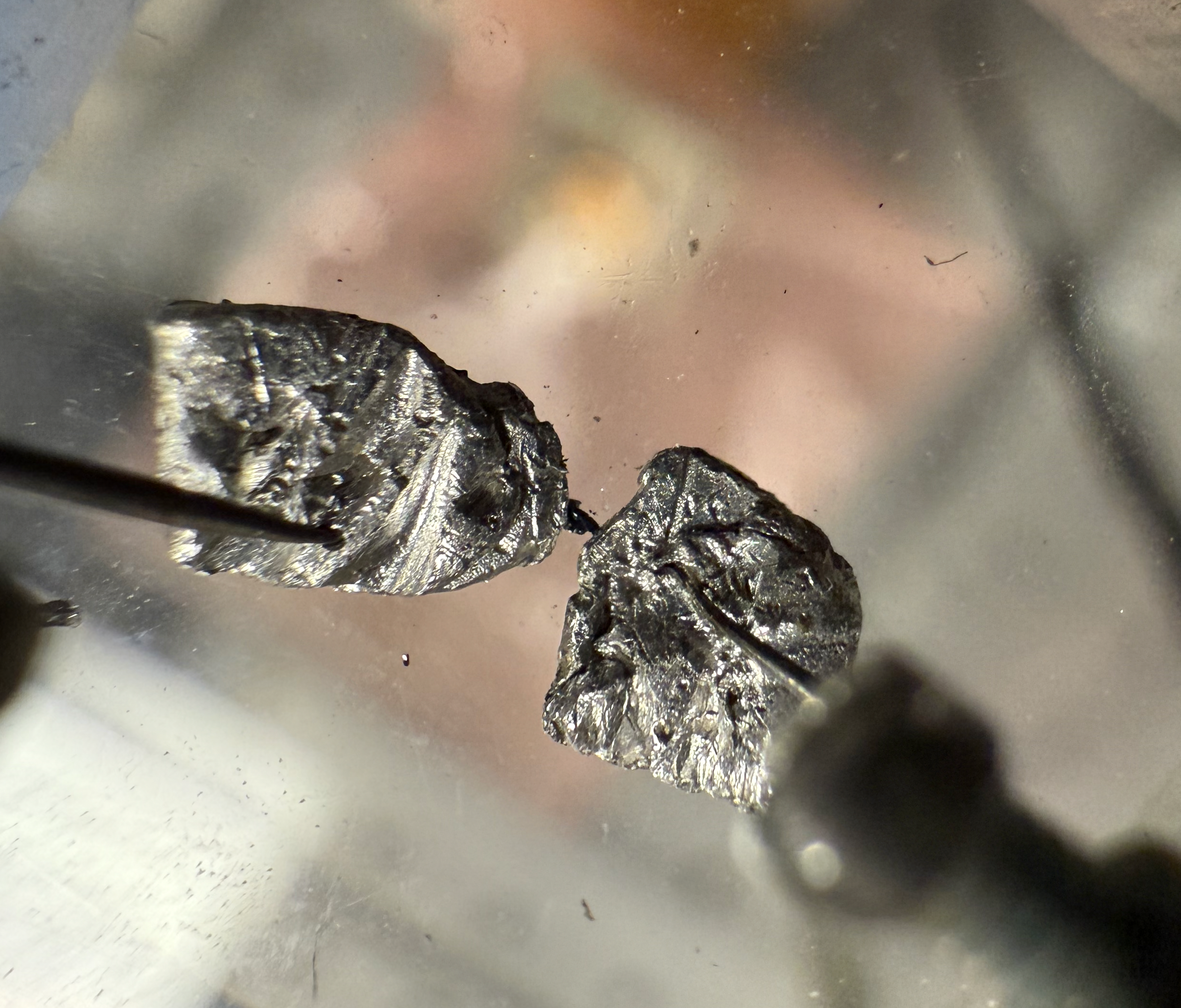}
    \caption{Photoconductive current measurement setup, where indium pads are used to make contact with BCP-Crystal.}
    \label{fig:s5}
\end{figure}

Another important step is to determine the laser power density by measuring the laser spot size. We measured the laser spot size by focusing the laser on a smooth Si wafer and subsequently increasing the laser power right until the microscope camera was saturated with the signal; at which point the central Airy disk and the first few diffraction rings of the laser spot were visible (\hyperref[fig:s4]{Figure S4}). We considered the spot size to be the size of the central Airy disk, which is a lower limit and assumes scattered light is not a significant source of excitation carrier. A Video of the laser spot size profile with increasing laser power up to signal saturation is in the GitHub repository \href{https://github.com/gideon116/BaCd2P2-aPIT-counterpart-of-GaAs-for-PVs/tree/main/laser_spot}{https://github.com/gideon116/BaCd2P2-aPIT-counterpart-of-GaAs-for-PVs/tree/main/laser\_spot}.

\section{Lifetime from Photoconductive Current}\label{sec:s4}

To measure the photoconductive current, we modulated a $780\ \mathrm{nm}$ laser diode at $227\  \mathrm{Hz}$ (square wave) using an SRS DS345 function generator. The laser diode was placed $1\  \mathrm{cm}$ above the samples and light from the laser diode diverges \(8^{\circ}\) and \(30^{\circ}\), resulting in an ellipse of area \(6.34\ \mathrm{mm^{2}}\). The maximum output power of the laser diode is $10.5\  \mathrm{mW}$. Therefore, the power density 1 cm away is \(0.164\ \mathrm{W/cm^{2}}\). The measured current was passed through an SR570 low-noise current preamplifier and a low-pass filter. An SR830 lock-in amplifier was used to isolate the current signal at $227\ \mathrm{Hz}$ and measure the photoconductive current. To make ohmic contacts, indium pads were pressed against BCP-Crystal, and indium was soldered onto GaAs-Wafer.

\subsection{Determining the Lifetime of BCP}

\begin{figure}
    \centering
    \includegraphics[width=0.5\linewidth]{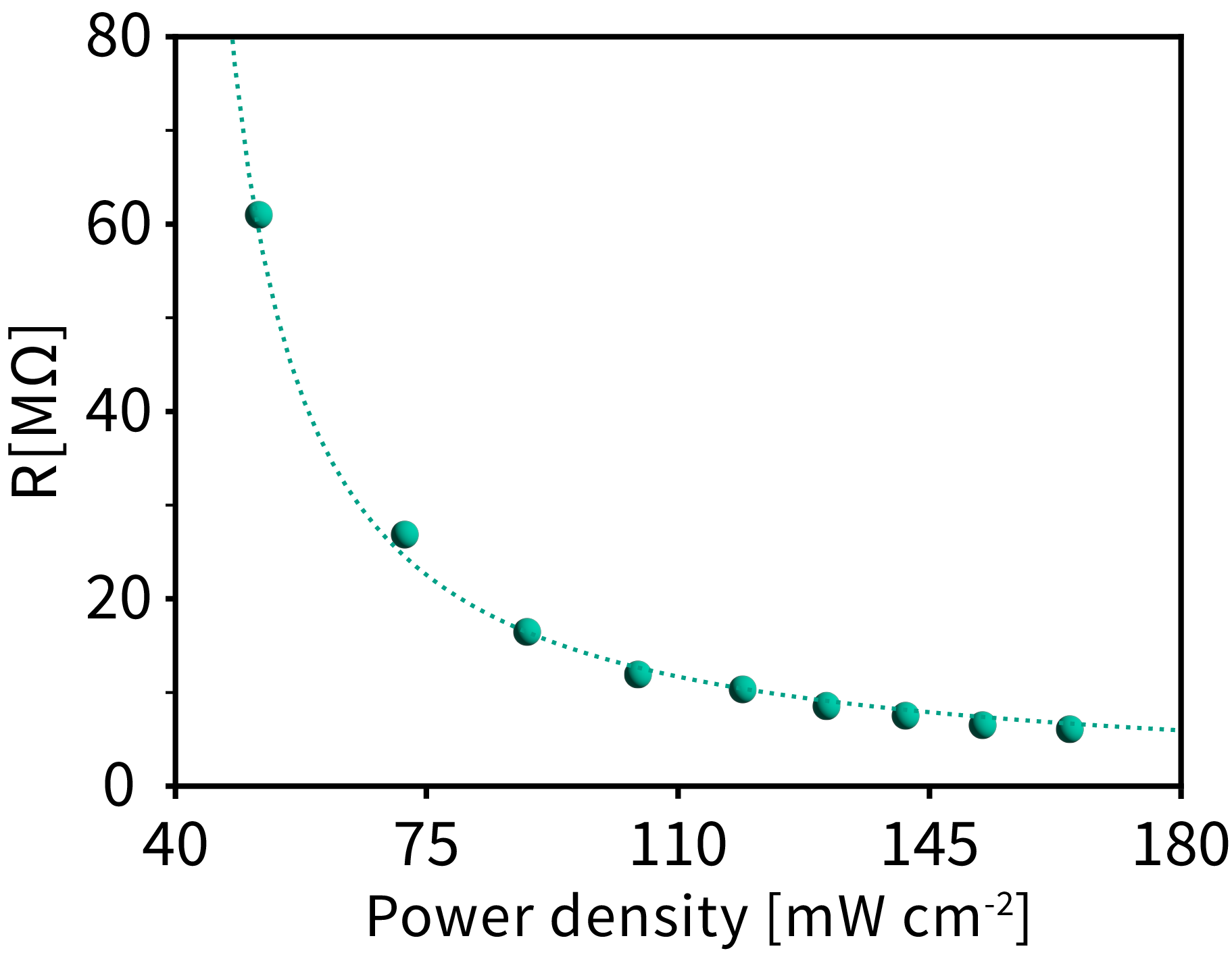}
    \caption{Change in $R_\mathrm{light}^\mathrm{meas}$ (measured resistance) as a function of incident power density. The dashed line represents fitting the data to \hyperref[eq:s38]{Equation S38}. The resulting $R_\mathrm{contact}$ (contact resistance) is $\approx0$.}
    \label{fig:s6}
\end{figure}

We can extract carrier lifetime from dark and photoconductive current measurements. Since we used a two-point probe configuration, we must correct for the contact resistance (\(\mathrm{R_{contact}}\)). Under illumination, the total resistance measured ($R_\mathrm{light}^\mathrm{meas}$) is the sum of the real sample resistance ($R_\mathrm{light}$) and twice the contact resistance, i.e.,  \(R_\mathrm{light}^\mathrm{meas} = R_\mathrm{light} + 2 \cdot \mathrm{R_{contact}}\). For a given photo generated carrier density, $\Delta n$, we know that:
\begin{equation}
R_\mathrm{light} = \frac{L}{(\mu_{n} + \mu_{p})\  (\mathrm{\Delta}n + n_{i}) \  q \ W\ D}
\label{eq:s37}
\end{equation}
where $\mu_{n}$ and $\mu_{p}$ are the electron and hole mobilities; $q$  is the elementary charge; $L$, $W$, and $D$ are the sample's length, width, and depth; and $n_i$ is the intrinsic carrier concentration. We can reasonably assume that \(\mathrm{R_{contact}}\) and the carrier mobilities remain the same under dark and illuminated measurements.  Furthermore, if we assume light is absorbed until $D$, and $\Delta n$ scales linearly with optical power (\(P_\mathrm{{opt}}\)), that is $\Delta n = K \cdot P_\mathrm{{opt}}$, then \(\mathrm{R_{contact}}\) can be extracted by fitting \(P_\mathrm{{opt}}\) vs \(R_\mathrm{light}^\mathrm{meas}\) (\hyperref[fig:s6]{Figure S6}) using the relation:
\begin{equation}
R_\mathrm{light}^\mathrm{meas} = \frac{L}{(\mu_{n} + \mu_{p}) \ (K \cdot P_\mathrm{{opt}} + n_{i}) \ q \ W \ D} + {2 \ R}_{contact}
\label{eq:s38}
\end{equation}
The optical power density was varied from \(0.053-0.133\ \mathrm{W/cm^{2}}\) on a BCP-Crystal of surface area \(L \times W = \ 1.69\ \mathrm{mm^{2}}\). Fitting \(P_\mathrm{{opt}}\) with \(R_\mathrm{light}^\mathrm{meas}\) using \hyperref[eq:s38]{Equation S38}, we find \(\mathrm{R_{contact}} \approx 0\), thus we can take $R_\mathrm{light} \approx R_\mathrm{light}^\mathrm{meas}$.

When \(\Delta n \gg n_i\), \hyperref[eq:s37]{Equation S37} rearranges to:
\begin{equation}
\mathrm{\Delta}n = \frac{L}{(\mu_{n} + \mu_{p}) \ R_\mathrm{light} \ q \ W \ D}
\label{eq:s39}
\end{equation}
Following absorption and diffusion of carriers, the generation rate of photoelectrons, $G_e$, is given by:
\begin{equation}
G_{e} = \frac{(\eta \  \frac{P_\mathrm{{opt}}}{hv} )}{L \ W \ D}
\label{eq:s40}
\end{equation}
where $\eta$ is the external quantum efficiency and $hv$ is the photon energy. Hence the carrier lifetime (\(\tau\)) is:
\begin{equation}
\tau = \frac{\Delta n}{G_e}
\label{eq:s41}
\end{equation}
We now consider another approach of obtaining $\tau$, where we do not neglect contact resistance.  To begin, we determine \(\mu_{n} + \mu_{p}\) using the resistances of pairs of varying illuminations and the dark resistance ($R_\mathrm{dark}^\mathrm{meas}$), where:
\begin{equation}
R_\mathrm{dark}^\mathrm{meas} = R_\mathrm{dark} + 2 \ \mathrm{R_{contact}} = \frac{L}{( \mu_{n} + \mu_{p}) \ n_i \ q \ W \ D} + 2 \ \mathrm{R_{contact}}
\label{eq:s42}
\end{equation}
Let \(x = \mu_{n} + \mu_{p}\) and \(C = \frac{q \ W \ D}{L}\ \). Now, when $\Delta n \gg n_i$, the resistance measured at illumination level  $light_1$ ($R_\mathrm{light_1}^\mathrm{meas}$) is given by: 
\begin{equation}
R_{\mathrm{light_1}}^\mathrm{meas} =R_\mathrm{light} + 2 \ \mathrm{R_{contact}}= \frac{1}{x \ \mathrm{\Delta}n_{1} \ C} + R_\mathrm{dark}^\mathrm{meas} - \frac{1}{x \ n_{i} \ C}
\label{eq:s43}
\end{equation}
Thus, it follows that:
\begin{equation}
\mathrm{\Delta}n_{1} = \frac{1}{\frac{1}{n_{i}} - x \ C \ \left( {R_\mathrm{dark}^\mathrm{meas} - R}_{\mathrm{light_1}}^\mathrm{meas} \right)}
\label{eq:s44}
\end{equation}
And comparing illuminations $\mathrm{light_1}$ and $\mathrm{light_2}$, measured using $P_\mathrm{{opt}}$ of $P_1$ and $P_2$ respectively, we find:
\begin{equation}
\frac{P_{1}}{P_{2}} = \frac{\mathrm{\Delta}n_{1}}{\mathrm{\Delta}n_{2}} = \frac{\frac{1}{n_{i}} - x \cdot C \cdot \left( {R_\mathrm{dark}^\mathrm{meas} - R}_{\mathrm{light_2}}^\mathrm{meas} \right)}{\frac{1}{n_{i}} - x \cdot C \cdot \left( {R_\mathrm{dark}^\mathrm{meas} - R}_{\mathrm{light_1}}^\mathrm{meas} \right)}
\label{eq:s45}
\end{equation}
So that:
\begin{equation}
x = \frac{1-\frac{P_1}{P_2}}{n_i \ C \ [(R_\mathrm{dark}^\mathrm{meas}-R_\mathrm{light_2}^\mathrm{meas})-\frac{P_1}{P_2}(R_\mathrm{dark}^\mathrm{meas}-R_\mathrm{light_1}^\mathrm{meas})]}
\label{eq:s46}
\end{equation}
Using $n_i \approx 2 \times 10^{7}\ \mathrm{cm^{- 3}}$ (as discussed in \hyperref[sec:s3]{SI Section 3}) we looked at eight measurements, each collected at varying illuminations, and all gave a value of $x = \mu_{n} + \mu_{p} = 4.9\ \mathrm{\frac{cm^{2}}{Vs}}$. Thus, we determine $\tau$ using:
\begin{equation}
\tau = \frac{\Delta n}{G_e} = \frac{L^2}{(\mu_{n} + \mu_{p}) \ R_\mathrm{light} \ q} \cdot \frac{hv}{\eta \ P_\mathrm{{opt}}}
\label{eq:s47}
\end{equation}
When both considering and neglecting contact resistance, we see that the lower limit of $\tau$ (when  $\eta = 1$) stabilizes at \(\tau \geq 300\ \mathrm{ns}\) under a power density of $0.15\  \mathrm{W/cm^2}$, confirming the assessment in \hyperref[fig:s6]{Figure S6}.

\subsection{Determining the Lifetime of GaAs}

In contract to BCP-Crystal, GaAs-Wafer was large enough $L = 10\ \mathrm{mm}$, $W = 10\ \mathrm{mm}$, $D = 0.5\ \mathrm{mm}$ for Hall effect measurements. The measurement showed that it has an Si dopant concentration of $1.6 \times 10^{17} \mathrm{cm^{- 3}}$ and electron mobility of $1740 \ \mathrm{\frac{cm^{2}}{Vs}}$. Using these values, a hole mobility of $\approx 400 \ \mathrm{\frac{cm^2}{Vs}}$ \cite{Hilsum1976SemiconductorsPhenomena}, and \hyperref[eq:s42]{Equation S42}, yields $\mathrm{R_{contact}} = 403\ \Omega$. Under illumination, we found $R_\mathrm{light}^\mathrm{meas} = 1.01 \times 10^{7}\ \Omega$. It then follows that $R_\mathrm{light} = R_\mathrm{light}^\mathrm{meas} - 2\ \mathrm{R_{contact}} \approx 1.01 \times 10^{7}\ \Omega$. Using \hyperref[eq:s41]{Equation S41} and the $\eta$ of GaAs at $780\  \mathrm{nm}$ ($eta \approx 0.9$), we find that $\tau \approx 5 \ \mathrm{ns}$.

\section{Computational Details}\label{sec:s5}

We performed first-principles density-functional theory (DFT) calculations using the Vienna Ab initio Simulation Package (VASP) and the HSE screened hybrid functional \cite{Kresse1993AbMetals, Kresse1996EfficiencySet,Heyd2003HybridPotential}. For the GaAs calculations, we used a mixing parameter of $0.28$. We optimized the primitive unit cell with an $8\times8\times8$ $\Gamma$-centered $\mathrm{k}$-point mesh and a force convergence criterion of 0.005 $\mathrm{eV}/$\AA. We used a plane-wave energy cutoff of $400\ \mathrm{eV}$  and $10^{-7}\ \mathrm{eV}$ as a total-energy convergence criterion for the electronic self-consistency loop. Using these parameters gives a band gap of $1.528\ \mathrm{eV}$, quite close to the accepted 0 K band gap of $1.519\ \mathrm{eV}$ GaAs \cite{Levinshtein1996HandbookParameters, Blakemore1982SemiconductingArsenide}.

\begin{figure}
    \centering
    \includegraphics[width=0.9\linewidth]{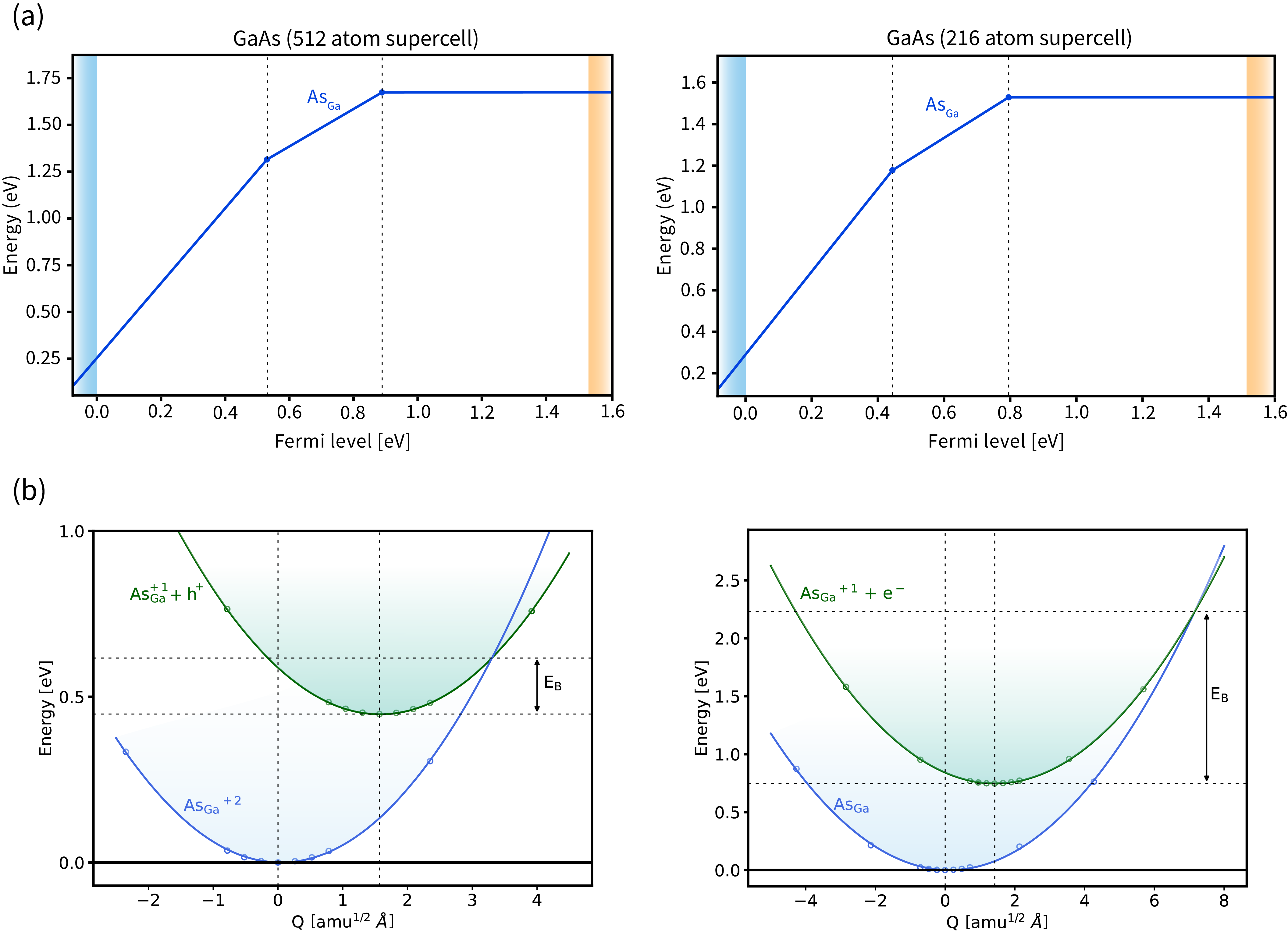}
    \caption{(a) The formation energy diagram of $\mathrm{As_{Ga}}$ under As-rich conditions using different sized supercells. Most screening calculations to identify the $\mathrm{As_{Ga}}$ transition corresponding to EL2 were performed using 216-atom supercells; here we present a convergence test of the 216-atom supercell against that of a 512-atom supercell. (b) The configuration coordinate diagrams for the $(+/2+)\ \mathrm{As_{Ga}}$ and $(+/2+)\ \mathrm{As_{Ga}}$ transitions, computed using a 216-atom supercell; $E_\mathrm{B}$ indicates the capture barrier.}
    \label{fig:s7}
\end{figure}

To model the $\mathrm{As}_\mathrm{Ga}$ antisite defect in GaAs, a 216-atom supercell was constructed, which is a $3 \times 3 \times 3$ repetition of the GaAs conventional unit cell, and a $\Gamma$-only $\mathrm{k}$-point mesh was used for Brillouin-zone integration. For the supercell containing $\mathrm{As}_\mathrm{Ga}$ in different charge states, all internal atomic positions were fully relaxed until the residual atomic force became less than 0.005 $\mathrm{eV}/$\AA. For analyzing the defect calculations, we used the Python toolkits PyCDT \cite{Broberg2018PyCDT:Insulators} and pydefect \cite{Kumagai2021InsightsCalculations}, and the total energy for charged defects was corrected using Kumagai's implementation of the Freysoldt-Neugebauer–Van de Walle correction scheme \cite{Freysoldt2009FullyCalculations,Kumagai2014Electrostatics-basedCalculations}. For the calculation of nonradiative carrier captures by $\mathrm{As}_\mathrm{Ga}$, we used Fermi's golden rule within the static coupling formalism and an one-dimensional configuration-coordinate model as implemented in the Nonrad code \cite{Turiansky2021Nonrad:Principles, Alkauskas2014First-principlesEmission}. We performed convergence tests using a 512-atom supercell, which is a $4 \times 4 \times 4$ repetition of the GaAs conventional unit cell. We found that the $(+/2+)\ \mathrm{As_{Ga}}$ transition level is $0.45\ \mathrm{eV}$ and $0.53\ \mathrm{eV}$ obtained from 216- and 512-atom supercell calculations, respectively, and that the semi-classical capture barrier ($E_\mathrm{B}$) is $0.20\ \mathrm{eV}$ and $0.24\ \mathrm{eV}$ respectively (\hyperref[fig:s7]{Figure S7}). On the other hand, we found the $(0/+)\ \mathrm{As_{Ga}}$ transition level to be $0.80\ \mathrm{eV}$ using a 216-atom supercell and $0.89\ \mathrm{eV}$ when using a 512-atom supercell, and conrrespondingly $E_\mathrm{B}=1.4\ \mathrm{eV}$. Both the transition level and $E_\mathrm{B}$ of $(+/2+)\ \mathrm{As_{Ga}}$ are better matches to the experimental values of EL2 \cite{Kaminska1987EL2GaAs, Martin1980KeyGaAs, Weber1982IdentificationGaAs, Kaminska1987EL2GaAs, Elliott1984IdentificationGaAs} than $(0/+)\ \mathrm{As_{Ga}}$. These results suggest that $(+/2+)\ \mathrm{As_{Ga}}$ is the EL2 defect. For BCP, we used our previous defect calculations in \cite{Yuan2024DiscoveryAbsorber}, which were produced using a mixing parameter of 0.25 and a $4\times 4\times 3$ supercell (240 atoms); the calculated band structure is reproduced in \hyperref[fig:s8]{Figure S8}.

% \begin{table}
%     \centering
%     \caption{Comparing the ionization level, $E_\mathrm{B}$, and $C_\mathrm{C}$ of the (+/+2) and (0/+) charge transition levels of $\mathrm{As_{Ga}}$ to identify the better match to experimentally observed values.\\}
%     \label{table:s4}
%     \begin{tabular}{lllcc}
%         \hline
%         Transition & Level ($\mathrm{eV}$)$^1$ & Captured & $E_\mathrm{B}$ (eV) & $C_\mathrm{C}$ ($\mathrm{cm}^3/\mathrm{s}$) \\
%         \hline
%         (+/+2) & 0.53 (0.45) & hole & 0.2 & $3 \times 10^{-10}$
%         \\
%         && electron & 2.2 & $2 \times 10^{-23}$
%         \\
%         \hline
%         (0/+) & 0.89 (0.80) & hole & 1.4 & $1 \times 10^{-20}$
%         \\
%         && electron & 1.4 & $3 \times 10^{-18}$
%          \\
%         \hline
%     \end{tabular}
%     \begin{tabular}{c}
%     $^1$in parenthesis are the 216 atom supercell values
%     \end{tabular}
% \end{table}

According to the principle of detailed balance, the Shockley-Read-Hall (SRH) nonradiative recombination rate ($R$) is given by \cite{Das2020WhatTheory,Shockley1952StatisticsElectrons}:
\begin{equation}
R = \frac{np - n_{i}^{2}}{\frac{1}{NC_{p}} \ \left( n + n_{1} \right) + \frac{1}{NC_{n}} \ (p + p_{1})},
\label{eq:s48}
\end{equation}
where $n_{i}$ is the intrinsic carrier concentration, and $N$ the defect concentration; $n$ and $p$ are the electron and hole concentrations, $C_{n}$ and $C_{p}$ are the electron and hole carrier capture coefficients, {\color{black} and $n_1$ and $p_1$ are the electron and hole concentrations when the Fermi level is at the defect level, respectively. We determine the total defect concentration, $N$, of a defect species $D$ with charge states $q$, the formation enthalpy (i.e., the formation energy) at a given set of atomic chemical potentials $\mu_i$ and Fermi level $E_F$ is \cite{VandeWalle2004First-principlesIII-nitrides, Freysoldt2014First-principlesSolids}
\begin{equation}
\Delta H_f(D^q; E_F,\mu_i) = E_{\mathrm{tot}}(D^q)-E_{\mathrm{tot}}(\mathrm{bulk}) -\sum_i n_i \mu_i + qE_F +E_{\mathrm{corr}}.
\label{eq:s49}
\end{equation}

It then follows that $\Delta G_f (D^q; E_F, \mu_i, T) = \Delta H_f(D^q; E_F,\mu_i) - T \Delta S(D^q; T)$, where $\Delta S$ is the total entropy, which includes the phonon (vibrational) entropy $S_\mathrm{ph}$, and $\Delta G_f$ is the Gibbs free energy of formation. Here, we neglect the $S_\mathrm{ph}$ contributions, an approximation common in computational works investigating point defects in solids \cite{Buckeridge2019EquilibriumEnergy}. Thus, assuming dilute, non-interacting defects, the concentration of charge state $q$ is \cite{Squires2023Py-sc-fermi:Calculations}:
\begin{equation}
[D^q](E_F,T) = \frac{N_{\mathrm{sites}}}{V}\; g_q\; \exp\!\left[-\frac{\Delta H_f(D^q;E_F,\mu_i)}{k_BT}\right],
\label{eq:s50}
\end{equation}
where $N_{\mathrm{sites}}$ is the number of available sites and $g_q$ is the degeneracy (together, $N_{\mathrm{sites}}$ and $g_q$ account for the configurational entropy), and $V$ is the cell volume. Thus, the total defect concentration is $N=[D]=\sum_q [D^q]$.

Using the host density of states $g(E)$ and the Fermi-Dirac distribution $f(E;E_F,T)=\left(1+e^{(E-E_F)/k_BT}\right)^{-1}$, the electron and hole concentrations are \cite{Squires2023Py-sc-fermi:Calculations}
\begin{equation}
n(E_F,T)=\int_{E_C}^{\infty} g(E)\  f(E;E_F,T)\  dE,
\qquad
p(E_F,T)=\int_{-\infty}^{E_V} g(E)\ \bigl[1-f(E;E_F,T)\bigr]\  dE.
\label{eq:s51}
\end{equation}

$E_F$ is obtained by solving the charge-neutrality condition \cite{Squires2023Py-sc-fermi:Calculations}
\begin{equation}
p(E_F,T)-n(E_F,T)+\sum_D\sum_q q\ [D^q](E_F,T)=0,
\label{eq:s52}
\end{equation}
which we solve iteratively to obtain the self-consistent $E_F$, then evaluate Equations \hyperref[eq:s50]{S50} and \hyperref[eq:s51]{S51} using this value.

\begin{figure}
    \centering
    \includegraphics[width=0.75\linewidth]{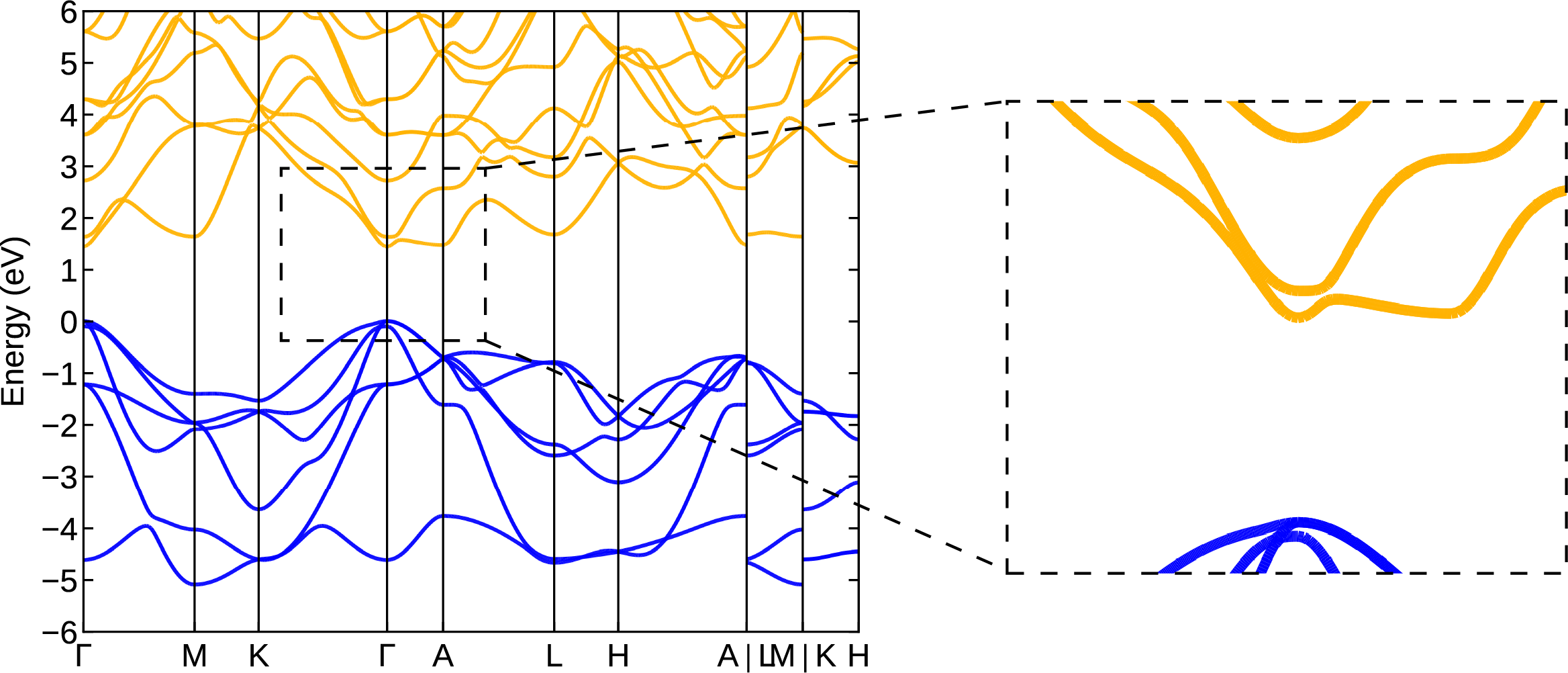}
    \caption{The band structure of BCP, as reported in our previous work \cite{Yuan2024DiscoveryAbsorber}, with a magnified view highlighting the two direct band edges responsible for the dual emission observed in PL.}
    \label{fig:s8}
\end{figure}

We are now in a position to calculate the SRH nonradiative recombination rates for the main intrinsic deep defects in BCP and GaAs, considering a typical photo-excited excess carrier density of $\mathrm{\Delta}n = 10^{14}\ \mathrm{cm}^{-3}$.}

\begin{figure}[!t]
    \centering
    {\color{black}
    \includegraphics[width=1.0\linewidth]{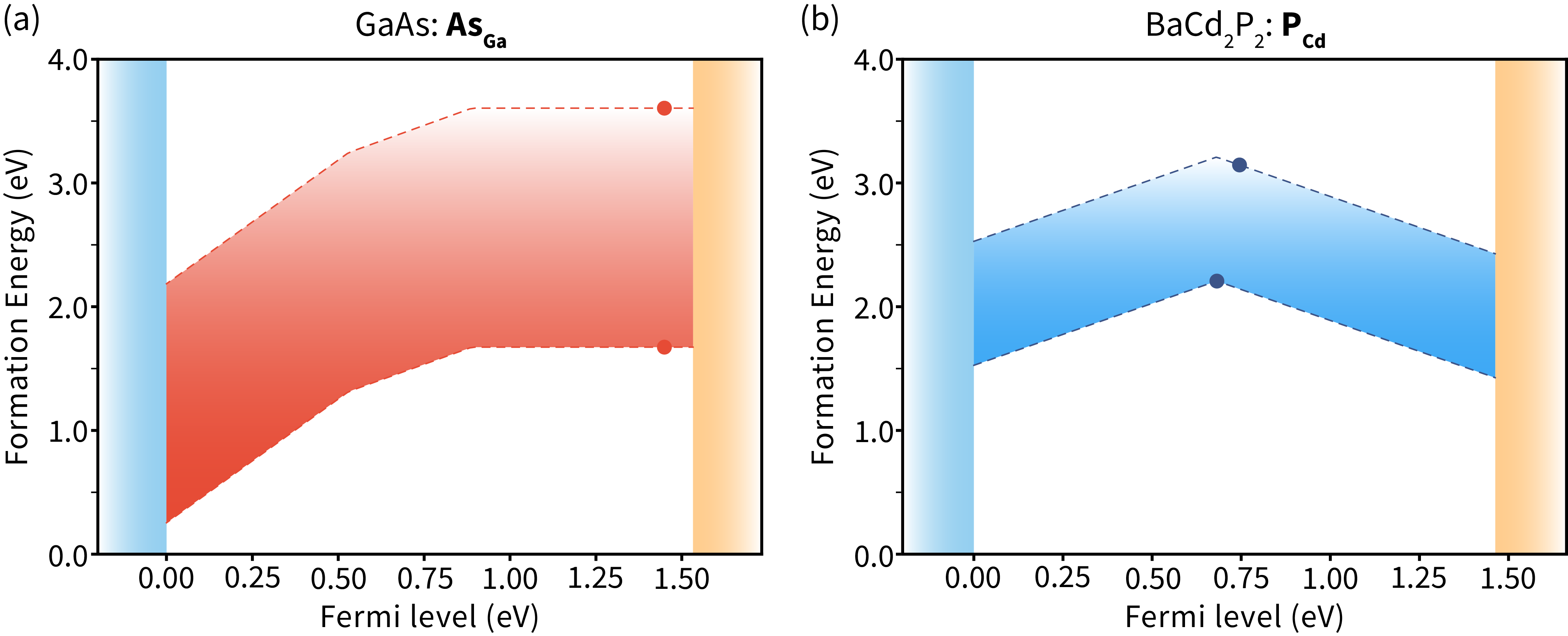}
    \caption{\label{fig:s9} The defect formation energy vs Fermi-level diagram for (a) $\mathrm{P_{Cd}}$ in BCP and (b) $\mathrm{As_{Ga}}$ in GaAs that has been Si-doped to $1.6\times 10^{17}\ \mathrm{cm^{-3}}$ across their entire stable chemical-potential region. The dashed lines represent the chemical potentials corresponding to the highest and lowest formation energies. The circles represent the self-consistent Fermi level at the respective chemical potentials. In (a), the upper bound for the formation energy of $\mathrm{As_{Ga}}$ is achieved under a chemical potential of As:-0.97 eV and Ga:0 eV, while the lower bound is achieved under As:0 eV and Ga:-0.97 eV. In (b), the upper bound of the formation energy of $\mathrm{P_{Cd}}$ is achieved under a chemical potential of Ba:-1.25 eV, Cd:-0.08 eV, and P:-0.95 eV, while the lower bound is achieved under Ba:-2.58 eV, Cd:-0.25 eV, and P:-0.12 eV.}}
\end{figure}

\textbf{GaAs}. Our n-type GaAs-Wafer has a dopant concentration of ${\sim10}^{17}\ \mathrm{cm}^{- 3}$. In this case, $p = \mathrm{\Delta}n \ll n$, which lies in the low-injection regime, and given that $E_\mathrm{B}$ is large for the capture of electron by $\mathrm{As_{Ga}}^{+2}$ \hyperref[eq:s48]{Equation S48} reduces to:
\begin{equation}
R = N \cdot \mathrm{\Delta}n \cdot C_{p}.
\label{eq:s53}
\end{equation}
For the $(+/2+)$ level $\mathrm{As_{Ga}}$, $C_{p} = 3.0 \times 10^{- 10}\ \mathrm{cm^{3}\ s^{-1}}$ at 300 K. The GaAs wafer used in our experiments was grown by the Liquid Encapsulated Czochralski method at a temperature of around 1510 K \cite{UniversityWafer2025GalliumArsenide}. We determined the defect concentration, Fermi-level position, and free carrier concentration under the charge-neutrality condition \cite{Buckeridge2019EquilibriumEnergy, Squires2023Py-sc-fermi:Calculations}. Defect concentrations are often calculated at synthesis temperatures, assuming most of the point defects formed during synthesis remain frozen in due to kinetic barriers \cite{Buckeridge2019EquilibriumEnergy}. However, this might not necessarily be the case. For instance, the experimentally reported concentration of EL2, accepted to be $\mathrm{As_{Ga}}$, is $\approx10^{16}\ \mathrm{cm}^{-3}$ \cite{Rudolph1999BulkOverview}. We see that this concentration is achieved at $\sim1160\ \mathrm{K}$ under As-rich conditions (the condition under which our GaAs sample was prepared) in the theoretical $\mathrm{As_{Ga}}$ defect concentration vs temperature curve in \hyperref[fig:5]{Figure 5a}. With this knowledge, we use $3/4\ \mathrm{th}$ of the synthesis temperature to calculate the defect concentrations for GaAs and BCP. Assuming the defect concentration at this temperature gets frozen, we find that, when $\mathrm{\Delta}n = 10^{14}\ \mathrm{cm}^{-3}$, the SRH recombination rate is $3.45\ \times 10^{20}\ \mathrm{cm^{-3}\ s^{-1}}$ for our As-rich grown GaAs.

\textbf{BCP}. Because BCP is intrinsic and $n_{i} \approx 10^{7}~\mathrm{cm}^{- 3}$, $\mathrm{\Delta}n = 10^{14}\ \mathrm{cm}^{-3}$ lies in the high-injection regime. Therefore, $n = p = \mathrm{\Delta}n \gg n_{i}$ and \hyperref[eq:s48]{Equation S48} reduces to:
\begin{equation}
R = N \cdot \mathrm{\Delta}n \cdot \frac{C_{p}{\cdot C}_{n}}{C_{p} + C_{n}} = N \cdot \mathrm{\Delta}n \cdot C_{tot}.
\label{eq:s54}
\end{equation}
The major nonradiative recombination pathway in BCP was found to be related to the $(-/+)$ transition of the $\mathrm{P}_\mathrm{Cd}$ antisite, which involves three charge states, $\mathrm{P}_\mathrm{Cd}^{+1}$, $\mathrm{P}_\mathrm{Cd}^{0}$, and $\mathrm{P}_\mathrm{Cd}^{-1}$. In this case, $C_{tot}$ can be obtained by \cite{Alkauskas2016RoleSemiconductors}:
\begin{equation}
C_{tot} = \frac{C_{p}^{0} + C_{n}^{0}}{1 + \frac{C_{p}^{0}}{C_{n}^{+}} + \frac{C_{n}^{0}}{C_{p}^{-}}},
\label{eq:s55}
\end{equation}
where $C_{p}^{0}$ represents the capture of a hole by the defect in the neutral charge state, $C_{n}^{+}$represents the capture of an electron by the defect in the $+1$ charge state, and $C_{p}^{-}$ is for hole capture by the $-1$ charge state. For the $(+/0)$ and $(0/-)$ levels of $\mathrm{P}_\mathrm{Cd}$, in our previous work \cite{Yuan2024DiscoveryAbsorber}, we computed the capture coefficient for all four carrier capture processes involved at 300 K: $C_{p}^{0} = 7.30 \times 10^{- 8}\ \mathrm{cm^{3}\ s^{-1}}$, $C_{n}^{0} = 3.87 \times 10^{-7}\ \mathrm{cm^{3}\ s^{-1}}$, $C_{n}^{+} = 1.33 \times 10^{- 6}\ \mathrm{cm^{3}\ s^{-1}}$, and $C_{p}^{-} = 6.39 \times 10^{- 7}\ \mathrm{cm^{3}\ s^{-1}}$. It then follows that $C_{tot} = 2.77\  \times \ 10^{- 7}~\mathrm{cm^{3}\ s^{-1}}$ at 300 K.

Our BCP samples were synthesized at 1000 K; therefore, in accordance with the foregoing discussion, we calculate the defect concentrations at 0.75 of the synthesis temperature, that is 750 K. The $\mathrm{P}_\mathrm{Cd}$ concentration ranges from $2.91 \times 10^{1}$ to $1.57 \times 10^{8}\ \mathrm{cm}^{-3}$ under varying elemental chemical potentials in the region where BCP is stable. It then follows that under the photo-excitation of $\mathrm{\Delta}n \approx 10^{14}\ \mathrm{cm^{-3}}$, the SRH recombination rate ranges from $8.05 \times 10^{8}$ to $4.36 \times 10^{15}\ \mathrm{cm^{-3}\ s^{-1}}$.

{\color{black} The detailed values used to calculate the SRH rates in BCP and GaAs are presented in Tables \hyperref[table:s4]{S4}, \hyperref[table:s5]{S5}, and \hyperref[table:s6]{S6} below}.

\begin{table}[H]
    \centering
    {\color{black} 
    \caption{The nonradiative carrier capture parameters of point defects in GaAs. Where the transition levels are referenced to the VBM. $E_\mathrm{B} = \mathrm{N/A}$ refers to when the excited and ground states do not cross in the configuration coordinate diagram.\\}
    \label{table:s4}
    \begin{tabular}{c c c c c c}
        \hline
        \hline
        Defect & Transition & $E_\mathrm{0}$ (eV) & Capture process & $E_\mathrm{B}$ (eV) & $C_{n/p}$ (cm$^{3}$/s) \\
        \hline
        As$_{\mathrm{Ga}}$  & (2+/+)    & 0.53 & h$^+$ & 0.2    & $3 \times 10^{-10}$ \\
                            &           &      & e$^-$ & 2.2    & $2 \times 10^{-23}$ \\
        \hline
        As$_{\mathrm{Ga}}$  & (+/0)     & 0.89 & h$^+$ & 1.4    & $1 \times 10^{-20}$ \\
                            &           &      & e$^-$ & 1.4    & $3 \times 10^{-18}$ \\
        \hline
        Ga$_{\mathrm{As}}$  & (-/-2)    & 0.69 & h$^+$ & N/A     & $1 \times 10^{-25}$ \\
                            &           &      & e$^-$ & 0.1    & $2 \times 10^{-9}$ \\
        \hline
        V$_{\mathrm{As}}$   & (+/0)     & 0.97 & h$^+$ & 5.36   & $<1\times 10^{-30}$ \\
                            &           &      & e$^-$ & N/A     & $<1\times 10^{-30}$ \\
        \hline
        V$_{\mathrm{Ga}}$   & (-2/-3)   & 0.83 & h$^+$ & 4.16   & $<1\times 10^{-30}$ \\
                            &           &      & e$^-$ & N/A     & $<1\times 10^{-30}$ \\
        \hline
        V$_{\mathrm{Ga}}$   & (-/-2)    & 0.55 & h$^+$ & -      & - \\
                            &           &      & e$^-$ & N/A     & $<1\times 10^{-30}$ \\
    \hline
    \end{tabular}}
\end{table}

\begin{table}[H]
    \centering
    {\color{black} 
    \caption{The nonradiative carrier capture parameters of point defects in BCP as presented in \cite{Yuan2024DiscoveryAbsorber}. Where the transition levels are referenced to the VBM.\\}
    \label{table:s5}
    \begin{tabular}{c c c c c c}
        \hline
        \hline
        Defect & Transition & $E_\mathrm{0}$ (eV) & Capture process & $E_\mathrm{B}$ (eV) & $C_{n/p}$ (cm$^{3}$/s) \\
        \hline
        P$_{\mathrm{Cd}}$   & (0/-) & 0.712 & h$^+$ & 0.12 & $6.39 \times 10^{-7}$ \\
                            &       &       & e$^-$ & 0.03 & $3.87 \times 10^{-7}$ \\
        \hline
        P$_{\mathrm{Cd}}$   & (+/0) & 0.784 & h$^+$ & 0.05 & $7.30 \times 10^{-8}$ \\
                            &       &       & e$^-$ & 0.07 & $1.33 \times 10^{-6}$ \\
        \hline
        Cd$_{\mathrm{P}}$   & (+/0) & 0.641 & h$^+$ & 0.51 & $3.65 \times 10^{-11}$ \\
                            &       &       & e$^-$ & 0.15 & $4.28 \times 10^{-9}$ \\
        \hline
        Cd$_{\mathrm{P}}$   & (+2/+)& 0.648 & h$^+$ & 0.92 & $2.58 \times 10^{-17}$ \\
                            &       &       & e$^-$ & 0.15 & $1.32 \times 10^{-8}$ \\
        \hline
        Cd$_{\mathrm{Ba}}$  & (0/-) & 0.889 & h$^+$ & 0.80 & - \\
                            &       &       & e$^-$ & 0.90 & - \\
        \hline
    \end{tabular}}
\end{table}

\begin{table}[H]
    \centering
    {\color{black} 
    \caption{The defect concentrations of $\mathrm{Ga_{As}}$ and $\mathrm{As_{Ga}}$ in GaAs and $\mathrm{P_{Cd}}$ in BCP at synthesis temperature ($T_\mathrm{syn}$) and at $^3/_4T_\mathrm{syn}$. Where $T_\mathrm{syn} = 1510\ \mathrm{K}$ for GaAs and $T_\mathrm{syn} = 1000\ \mathrm{K}$ for BCP.\\}
    \label{table:s6}
    \begin{tabular}{l | l | l | l | l }
        \hline
        \hline
        Material    & Defect                & Chem Pot  & N at $T_\mathrm{syn}$ ($\mathrm{cm^{-3}}$) & N at  $^{3}/_{4}T_\mathrm{syn}$  ($\mathrm{cm^{-3}}$)  \\
        \hline
        GaAs        & $\mathrm{Ga_{As}}$    & Ga rich   & $1.58\times 10^{10}$  & $5.68\times 10^{6}$           \\
                    &                       & As rich   & $5.65\times 10^{4}$   & $2.31\times 10^{-2}$          \\
                    \cline{2-5}
                    & $\mathrm{As_{Ga}}$    & Ga rich   & $1.69\times 10^{11}$  & $4.69\times 10^{7}$           \\
                    &                       & As rich   & $4.58\times 10^{17}$  & $1.15\times 10^{16}$          \\
        \hline
        BCP         & $\mathrm{P_{Cd}}$     & P rich    & $6.71\times 10^{11}$  & $1.57\times 10^{8}$           \\ 
                    &                       & Cd rich   & $5.98\times 10^{6}$   & $2.90\times 10^{1}$           \\ 
        \hline
    \end{tabular}}
\end{table}

% \begin{figure}[!t]
    % \centering
    % \includegraphics[width=1.0\linewidth]{Figure5.png}
    % \caption{\label{fig:5}Calculated SRH nonradiative recombination rate for (a) BCP and (b) GaAs across their entire stable chemical-potential region; the regions with the red gradient represent the BCP and GaAs phases, respectively. The axes represent the chemical potential ($\mu$) of the constituent elements. The main deep defect for BCP is $\mathrm{P_{Cd}}$ and its deep (0/+) and (0/$-$) transitions are considered \cite{Yuan2024DiscoveryAbsorber}, while the main deep defect for GaAs is $\mathrm{As_{Ga}}$ and its deep (+/+2) transition is considered \cite{Kaminska1987EL2GaAs, Kaminska1993ChapterGaAs, Kaminska1985IdentificationDefect, SammyKayali1997GaAsApplications, Chadi1988MetastabilityGaAs, Dabrowski1988TheoreticalEL2, Wampler2015TemperatureArsenide}. The competing phases are labeled on the BCP chemical potential diagram. Note that the stable chemical potential range for GaAs is a line, and here this line is slightly expanded to a parallelogram for illustration purposes.}
% \end{figure}

% \begingroup
%   \sloppy
%   \printbibliography[
%     title={Supporting Information References},
%     section=2]
% \endgroup

\begingroup
  \sloppy
  \printbibliography[
    title={References}]
\endgroup

\end{refsection}

\end{document}